\documentclass[journal,twoside]{IEEEtran}

\usepackage{cite}

\usepackage[latin1]{inputenc} 			% Required for inputting international characters
\usepackage{color,soul}					% Highlighting text
\usepackage{array}
\usepackage{xtab,afterpage}
\usepackage{amsmath}

\ifCLASSINFOpdf
  \usepackage[pdftex]{graphicx}
   \graphicspath{{../figures/}}
\else
\fi

\newcolumntype{C}[1]{>{\centering\let\newline\\\arraybackslash\hspace{0pt}}m{#1}}
\usepackage{multirow}
\usepackage{xfrac}
\usepackage{url}
\usepackage{caption,float}
\usepackage{subcaption}

\usepackage{multicol}

\begin{document}

%\title{Who is the director of this film?\\Shot analysis of art films}
%\title{Who is the director of this film?\\Art movie authorship based on shot analysis}
\title{Who is the director of this movie?\\Automatic style recognition based on shot features}

\author{Michele Svanera*\thanks{Corresponding author. Michele.Svanera@glasgow.ac.uk},
	Mattia Savardi,
	Alberto Signoroni, %~\IEEEmembership{Member,~IEEE,}
	Andr\'as B\'alint Kov\'acs,
	Sergio Benini%~\IEEEmembership{Member,~IEEE}
\thanks{Authors are with Department of Information Engineering at University of Brescia (Italy), except for Andr\'as B\'alint Kov\'acs who is with Dept. of Film Studies at ELTE University (Bupadest, Hungary).}
\thanks{Manuscript received xxx xx, xxxx; revised xxx xx, xxxx.}
}

%\markboth{IEEE TRANSACTIONS ON MULTIMEDIA,~Vol.~xx, No.~x, Month~xxxx}
%{Svanera \MakeLowercase{\textit{et al.}}: Who is the director of this movie? Art movie authorship based on shot analysis}

\maketitle

%%%%%%%%%%%%%%%%%%%%%%%%%%%%%%%%%%%%%%%%%%%%%%%%%%%%%%
\begin{abstract}
We show how low-level formal features, such as \emph{shot duration}, meant as length of camera takes, and \emph{shot scale}, i.e. the distance between the camera and the subject, are distinctive of a director's style in art movies.
So far such features were thought of not having enough varieties to become distinctive of an author.
However our investigation on the full filmographies of six different authors (\emph{Scorsese, Godard, Tarr, Fellini, Antonioni}, and \emph{Bergman}) for a total number of 120 movies analysed second by second, confirms that these shot-related features do not appear as random patterns in movies from the same director.
For feature extraction we adopt methods based on both conventional and deep learning techniques.
Our findings suggest that feature sequential patterns, i.e. how features evolve in time, are at least as important as the related feature distributions.
To the best of our knowledge this is the first study dealing with automatic attribution of movie authorship, which opens up interesting lines of cross-disciplinary research on the impact of style on the aesthetic and emotional effects on the viewers.

\end{abstract}

%%%%%%%%%%%%%%%%%%%%%%%%%%%%%%%%%%%%%%%%%%%%%%%%%%%%%%

%%%%%%%%%%%%%%%%%%%%%%%%%%%%%%%%%%%%%%%%%%%%%%%%%%%%%%
\section{Introduction}

\begin{figure}[!ht]
    \begin{multicols}{2}
     \centering
      \begin{subfigure}{\linewidth}
      \centering
        \includegraphics[width=0.8\linewidth]{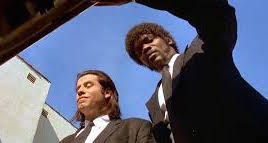}
        \caption{Pulp Fiction}
      \end{subfigure}
      \hfill
     \par
      \begin{subfigure}{\linewidth}
      \centering
        \includegraphics[width=0.8\linewidth]{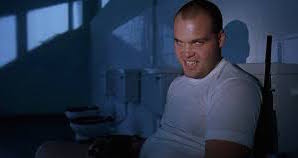}
        \caption{Full Metal Jacket}
      \end{subfigure}
      \par
      \begin{subfigure}{\linewidth}
      \centering
        \includegraphics[width=0.8\linewidth]{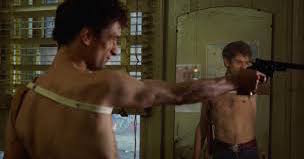}
        \caption{Taxi Driver}
      \end{subfigure}
      \par
      \begin{subfigure}{\linewidth}
      \centering
        \includegraphics[width=0.8\linewidth]{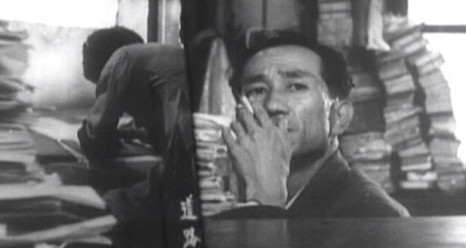}
        \caption{Ikiru}
      \end{subfigure}
      \par
      \begin{subfigure}{\linewidth}
      \centering
        \includegraphics[width=0.8\linewidth]{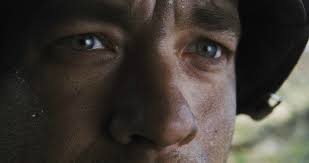}
        \caption{Saving Private Ryan}
      \end{subfigure}
      \par
      \begin{subfigure}{\linewidth}
      \centering
        \includegraphics[width=0.8\linewidth]{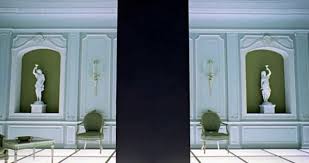}
        \caption{2001: A Space Odissey}
      \end{subfigure}
    \end{multicols}
 \caption{Some stylistic features make the author recognizable even when appearing once \cite{buzzfeed}: (a) the trunk shots (Tarantino), (b) the bathroom scenes (Kubrick), (c) the mirror shots (Scorsese). 
Other low-level features are indicative of the style when they are frequently used: (d) wipe effects (Kurosawa), (e) extreme close-ups (Spielberg), (f) symmetric shots (Kubrick).}
\label{fig:visual-formal}
\end{figure}

\IEEEPARstart{T}{he} impression that film authors have very different and recognizable styles is very common.
Some high-level stylistic features are capable of characterizing individual authors even when they appear only once in a film: if there is a shot taken from inside a car trunk, you are probably watching a Tarantino's movie (Figure~\ref{fig:visual-formal}(a)); Kubrick is famous for his pivotal scenes in bathrooms (Figure~\ref{fig:visual-formal}(b)); Scorsese often depicts characters in front of mirrors (Figure~\ref{fig:visual-formal}(c)).

However, as suggested by Salt \cite{S-06}, the obvious approach in searching for individual characteristics in the formal side of a director should consider those features that are most directly under his control\footnote{The individuals with the most control over a film are the \emph{director}, the \emph{cinematographer}, who manages the camera, lenses, etc., and the \emph{editor}, whose efforts are on pace. 
As in the infancy of motion pictures, when they were often one individual, we use \emph{director} or \emph{author}, to refer to the joint role of the three.}: \emph{shot duration}; \emph{shot scale}, i.e. the distance from the camera to the subject; \emph{shot transitions} such as cut, fades, wipes; \emph{camera movement} such as pan, tilt, zooms; \emph{camera angle} such as high-angle, low-angle, bird's eye, etc.

These low-level features represent the fundamental options of film technique.
They were thought in the past of not having enough varieties to become distinctive features of an author's style: a ``wipe effect'' by Kurosawa (Figure~\ref{fig:visual-formal}(d)), or an ``extreme close-up'' by Spielberg (Figure~\ref{fig:visual-formal}(e)), or a ``one-point perspective'' by Kubrick (Figure~\ref{fig:visual-formal}(f)) cannot be distinctive features of a style, if taken alone. 
%They are rather simple choices in a given moment of a film. 
But if a filmmaker adopts one feature conspicuously frequently when most of his colleagues act differently, this feature becomes distinctive not by its presence, but by the frequency of its occurrence. 

As a confirmation, recent cinema studies show that statistical analysis of shot features may reveal recurrent patterns in an author's work. 
For example, in \cite{Ko14} Kov\'acs disclosed systematic patterns of shot scale distributions in films by Antonioni.

Our hypothesis is that the statistical distribution and temporal pattern of a basic set of formal characteristics, such as shot duration and scale, act as a fingerprint of a specific director. 
The research questions we challenge in this work are therefore:
\begin{itemize}
\item Q1: Are shot duration and shot scale a sufficient feature set to recognise a director?
\item Q2: Which is the most distinctive feature for a director?
\item Q3: Is it possible to robustly determine the author of a movie using automatically computed shot features?
\end{itemize}

%
%%%%%%%%%%%%%%%%%%%%%%%%%%%%%%%%%%%%%%%%%%%
%\section{Previous work and contributions}

\subsection{Related work}
Statistical analysis of aesthetic forms goes back to the 1930s \cite{W40}. 
Characterizing individual, group or period styles by quantitative analysis of recurrent formal elements has been a currency especially in literature and music over the past decades. 
Conversely research of quantitative analysis of film style or automatic recognition of moving pictures' stylistic features has not developed in a similar pace due to the medium's increased sensory complexity as compared to other forms of art. 
In film studies this method was introduced by Salt in the 1970s \cite{S-06}, who started to annotate the presence of formal features, but this procedure has not become mainstream. 

Considering low-level features, the visual dimension is probably the most regarded by authors while planning the emotional impact of a film.
Automatic analysis often employs conventional MPEG7 color features which include color structure, layout, dominant color and scalable one \cite{MOVY01}.
Since previous work proved that saturation and color difference are crucial for mood elicitation in subjects \cite{VM94} also average saturation as well as variance play an important part.
Other visual features with a conspicuous role in media are color energy and lighting key \cite{WC06}.
Color energy depends on saturation, brightness and the occupied area, and it measures the perceptual strength of color.
Lighting keys, originally proposed in \cite{WC06}, describe two aesthetic techniques which are often employed: \emph{chiaroscuro} (i.e. high contrast between light and dark areas), and \emph{flat lighting} (i.e. low contrast). 
Last, the light spectral composition used during shooting, called illuminant (e.g. computed as in \cite{WGG07} and \cite{CBML09}), is important for assigning connotation to a scene.

Also motion dynamics are carefully planned by directors who rely on camera and object motion to transmit sensations of dynamism or quiet.
Several algorithms solve the problem of extracting camera motion parameters from image sequences to assess pan, tilt, zoom, roll, and horizontal/vertical tracking. 
Since the early 2000s many studies demonstrate that camera motion can be reliably estimated by processing data from the compressed stream \cite{TSK00,EST04}. 
A camera motion classification is for example used in \cite{BMS14} for distinguishing cinematographic shots into these classes: Aerial, Bird-eye, Crane, Dolly, Establishing, Pan, Tilt, and Zoom. 
Motion activity maps (MAM) as in \cite{ZGZ-02} keep into account both global motion and its spatial distribution in the scene.
Similarly to other motion features (e.g. MPEG-7 motion activity \cite{JD-01}) MAMs consider the overall intensity of motion activity, without distinguishing between camera and object motion.
In \cite{RSS05} motion content, computed in terms of structural tensor, is used to evaluate the \emph{visual disturbance} of scenes, with higher values for action films and less expected for dramatic or romantic ones.

%%%

Other variables under the director's control are depth of field, shot length, and scale.
Depth of field is the range of acceptable sharpness when a lens focuses on a subject at a distance.
Recovering it from single images is a challenging problem, addressed for example in \cite{ZS11} by estimating the amount of spatially varying defocus blur at edge locations. 

Shot duration and scale are also important formal aspects.
Shot boundary detection, with the purpose of temporal segmentation of videos into basic temporal units, has been a fundamental step of content analysis since early years \cite{H02} to more recent days \cite{SOD10}.
It is well known how shot duration influences audience perception: longer durations relax connotation, whereas shorter shots contribute to a faster pace \cite{Choros09}.

Regarding shot scale, varying camera distance is a common directing rule used to adjust the relative emphasis between the filmed subject and the surrounding scene \cite{WC09}.
This affects the emotional involvement of the audience \cite{A-91} and the identification process of viewers with the movie characters. 
In \cite{BSA16} we propose an automatic recognition method for shot scale by using intrinsic characteristics of shots, while in \cite{CBL11} we show how patterns of shot scale induce affective reactions in the audience, especially on the arousal dimension.

Previously mentioned features prove to be effective in a number of applications, ranging from video indexing \cite{SW05} to retrieval \cite{NH01}, affective analysis \cite{HX05} for video recommendation \cite{BCL11, CBL13a}, video abstraction and summarization \cite{NMZ05, MHL05}, just to name a few.
Restricting the analysis to movies, computable features inspired by cinematic principles are good indicators of different corpora of films based on periods, genres (such as drama, comedy, noir, etc. \cite{RSS05, ZHK10}), authors, or narratives.
In general detecting single films as belonging to these groups is important for various aims, such as movie recommendation \cite{MYL11}, film-therapy \cite{DH00}, or promotion \cite{DCC11}.

When dealing with the problem of author attribution, one attempt to recognize authorship through the use of visual characteristics is described in \cite{NHP15}, where authors train a neural network for attributing previously unseen painting to the correct author with a fair accuracy.
Automatic recognition of visual features in fine arts appears also with some convincing results in \cite{GEB15} which proposes a method to separate (and recombine) content and style present in arbitrary art images. 

To the best of our knowledge, no previous research tackled the problem of automatic attribution of movie authorship.

\subsection{Paper contributions}

This work is the first attempt to employ low-level formal features from movies as indicators of the idiosyncratic style of a director.
Following the intuition by Kov\'acs, who found in \cite{Ko14} systematic patterns of shot-based features in films by Antonioni, we prove the feasibility of author style recognition on a large, and probably unique, database of art movies.

\begin{figure}[!ht]
     \centering
      \begin{subfigure}{0.46\columnwidth}
        \includegraphics[width=\columnwidth]{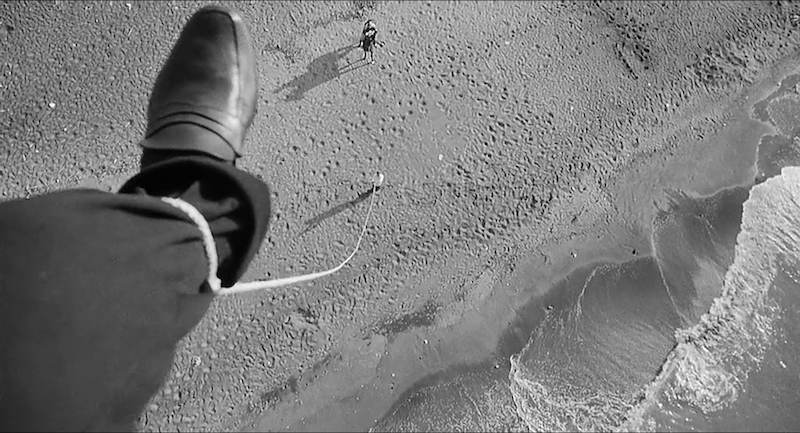}
        \caption{\emph{8\sfrac{1}{2}}}
      \end{subfigure}
      %\par
      \begin{subfigure}{0.49\columnwidth}
        \includegraphics[width=\columnwidth]{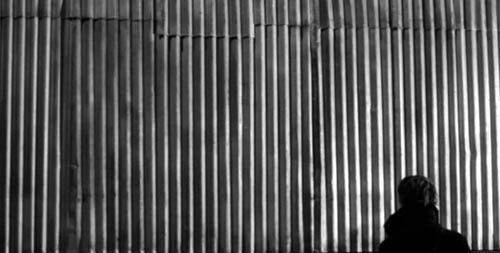}
        \caption{\emph{Werckmeister Harmóniák}}
      \end{subfigure}
      %\par
      \begin{subfigure}{0.46\columnwidth}
        \includegraphics[width=\columnwidth]{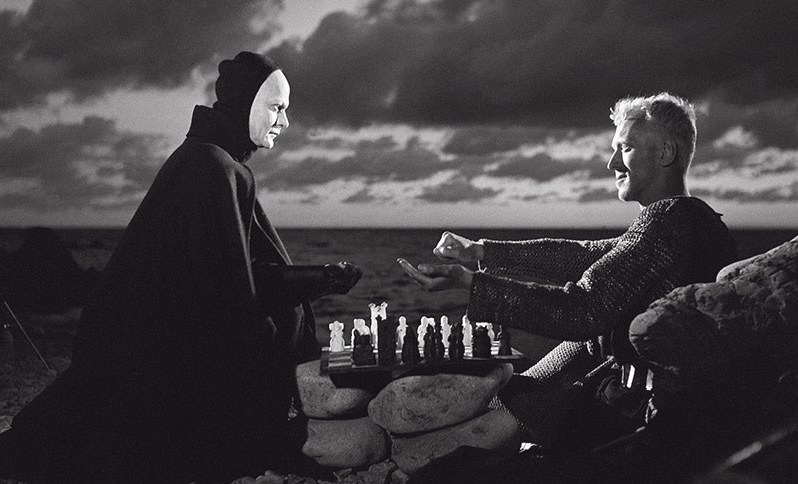}
        \caption{\emph{Det Sjunde Inseglet}}
      \end{subfigure}
     % \par
      \begin{subfigure}{0.49\columnwidth}
        \includegraphics[width=\columnwidth]{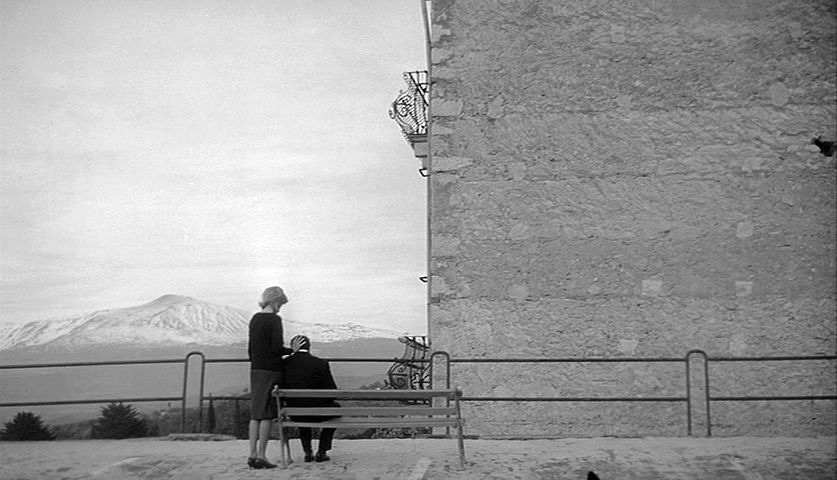}
        \caption{\emph{L'Avventura}}
      \end{subfigure}
 \caption{Examples of art movie content taken from a) \emph{Fellini, 1963}; b) \emph{Tarr, 2000}; c) \emph{Bergman, 1957}; d) \emph{Antonioni, 1960}.}
  \label{fig:art-movies}
\end{figure}

The dataset includes the almost complete filmography by six different directors whose styles are consensually considered highly unique and distinguishable in film historiography of author cinema: Michelangelo Antonioni, Ingmar Bergman, Federico Fellini, Jean-Luc Godard, Martin Scorsese, and B\'ela Tarr, for a total number of 120 movies.
The challenges posed by such artistic video content lie in the variety of experimental aesthetic situations, the richness of the scene composition, and the presence of unconventional or highly symbolic content \cite{wiki-art-film} (see examples in Figure~\ref{fig:art-movies}).

To answer the research questions we pose, we first perform a statistical analysis on manually annotated shot durations and scales.
These features are investigated not only in terms of their relative quantities, but also considering their sequential patterns. 
Being able to accurately model these temporal dependencies will be crucial for capturing the distinctive elements which are direct expressions of the directors' creativity.

Verified the possibility to recognize the director by relying on manually annotated features, we then repeat the classification by fully automatic means, with no drop in performance.
Since different studies \cite{CBD11, C16a} suggest a severe correlation between features and the movie period, we highlight how for the considered movies the feature set is director specific and not dependent from the stylistic period.
As a last contribution, we propose a method for shot scale recognition based on deep neural networks which outperforms existing state-of-the-art.

This document is organised as follows.
After illustrating the feature set in Section \ref{sec:features} and the movie database in Section \ref{sec:database}, we perform author recognition based on manual annotations in Section \ref{sec:exp1}.
This analysis, which answers Q1, is carried out by using a subset of $77$ movies for which we have the ground-truth (GT) of both shot duration and shot scale.
On the same film subset we carry out the analysis in Section \ref{sec:insights} to understand the relevance of each feature in the classification, the inter-feature correlation, their dependency with respect to the year of movie production, and the most distinctive feature for each director (Q2).
Finally, to tackle Q3, we propose an automatic framework based on both conventional computer vision and deep learning techniques to compute shot features (in Section \ref{sec:automatic_shot_analysis}) by which performing automatic authorship (in Section \ref{sec:exp2}).
After discussing obtained results in Section \ref{sec:discuss}, conclusions are finally drawn in Section \ref{sec:conclusion}.

%%%%%%%%%%%%%%%%%%%%%%%%%%%%%%%%%%%%%%%%%%%%%%%%%%%%%%%%%%%%%%%%%%%%%%%%%%%%%%%%%%%%%%%%%%%%%%%%%%%%%%%%%%%%%
\section{Feature sets}
\label{sec:features}

%We describe movies in terms of shot \textit{duration} and \emph{scale}.

%%%%%%%%%%%%%%%%%%%%%%%%%%%%%%%%%%%%%%%%%%%%%%%
\subsection{Shot duration features}

A shot is a series of consecutive frames from a single camera take showing a continuous action in time and space. 
As detailed in \cite{CBD11} and \cite{S12} an accurate use of shot duration, and its coupling with motion, allow the filmmaker to control the viewers' eye movements and modulate their attention.

Shots are here categorised into seven classes according to their duration $t_d$ (in seconds), as in Table~\ref{tab:classes-shot-duration}.
%
%\[
%\begin{cases}
%\text{\small Very Short (VS)} & if\:\: t_d\in(0,2) [s]\\
%\text{\small Short (S)} & if\:\:  t_d\in[2,4.5)[s]\\
%\text{\small Short Medium (SM)} & if\:\:   t_d\in[4.5,7)[s]\\
%\text{\small Medium (M)} & if\:\:   t_d\in[7,10)[s]\\
%\text{\small Medium Long (ML)} & if\:\:  t_d\in[10,22.5)[s]\\
%\text{\small Long (L)} & if\:\:  t_d\in[22.5,40)[s]\\
%\text{\small Very Long (VL)} & if\:\:  t_d\geq40 [s]
%\end{cases}
%\label{eq:sdd}
%\]

\begin{table}[ht]
\centering
      \caption{Classes of shot duration.}
 \begin{tabular}{ l | l   }
    \hline
    \textbf{Class} & \textbf{duration [s] }  \\ \hline  \hline
\text{\small Very Short (VS)} &  $t_d\in(0,2) $\\
\text{\small Short (S)} &   $t_d\in[2,4.5)$\\
\text{\small Short Medium (SM)} &    $t_d\in[4.5,7)$\\
\text{\small Medium (M)} &   $t_d\in[7,10)$\\
\text{\small Medium Long (ML)} &  $t_d\in[10,22.5)$\\
\text{\small Long (L)} &  $t_d\in[22.5,40)$\\
\text{\small Very Long (VL)} & $t_d\geq40$\\
\hline
  \end{tabular}
    \label{tab:classes-shot-duration}
  \end{table}

This subdivision is obtained by fitting a log-normal probability density function on GT shot duration data, thus achieving a finer granularity in proximity of the mode. 
To account for the evolution of pace along years, we consider studies in \cite{C16, C16a}, which report that the average shot duration for a sample of movies from 1960 to 1985 is 7.0 s, while in contemporary movies decreases to 4.3.
As an example of the ability of the proposed subdivision to describe shot durations in movies from different \'epoques and pace, inspect Figure~\ref{fig:DDistr_comparison} for shot durations from \textit{Le notti di Cabiria} (Fellini, 1957) and the more recent \textit{Gangs of New York} (Scorsese, 2002). 

%As an example of the ability of the proposed subdivision to describe shot durations in movies from different \'epoques and pace, we show in Figure~\ref{fig:SDD_example}(a) the distribution of shot duration for \textit{Le notti di Cabiria} (Fellini, 1957) and the more recent \textit{Gangs of New York} (Scorsese, 2002) in Figure~\ref{fig:SDD_example}(b).
%
%\begin{figure}[!ht]
%     \centering
%      \begin{subfigure}{0.49\columnwidth}
%        \includegraphics[width=\columnwidth]{1957_Fellini_-_Le_notti_di_Cabiria_GTonly.png}
%        \caption{\emph{Le notti di Cabiria}}
%      \end{subfigure}
%      %\par
%      \begin{subfigure}{0.49\columnwidth}
%        \includegraphics[width=\columnwidth]{2002_Scorsese_-_Gangs_of_New_York_GT_Only.png}
%        \caption{\emph{Gangs of New York}}
%      \end{subfigure}
%      %\par
%\caption{Distributions of shot duration: (a) \textit{Le notti di Cabiria (Fellini, 1957)}; (b) \textit{Gangs of New York (Scorsese, 2002)}.}
%  \label{fig:SDD_example}
%\end{figure}

Shot duration is here described by a 56-dimensional feature vector obtained by concatenating:
\begin{itemize} 
\item the shot Duration Distribution (\emph{DDistr}) over the seven classes (7-dimensions);
\item the shot Duration Transition matrix (\emph{DTrans}), which describes the probability transition in time of going from any class to any other class (49-dimensions).
\end{itemize}

\subsection{Shot scale features}
\label{ssec:exp-scale}
Shot scale, i.e. the relative distance of the camera from the main object of the film image, is one of the main ingredients of a film, with both stylistics \cite{Ko14} and narrative functions \cite{BKP16}. 
In particular the usage of the shot scale impacts on the narrative engagement of the viewer by empowering empathic emotions, and on the viewer's ability of attributing mental states to movie characters (the so called ``theory of mind'') \cite{BKP16}. 

\begin{figure}[h]
           \centering
       \includegraphics[width=1.0\columnwidth]{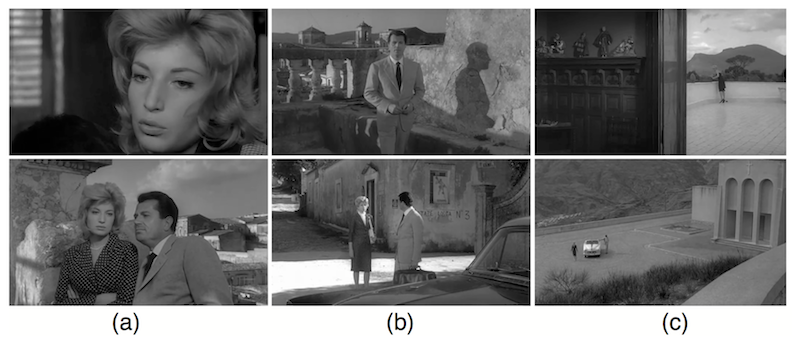}
       \caption{Shot scales: (a) Close, (b) Medium, and (c) Long shots, from \emph{L'Avventura (Antonioni, 1960)}.}
       \label{fig:types}
       \end{figure}

Although possible distances between camera and the filmed subject are infinite, in cinema studies the shot scale is usually mapped into seven categories: \emph{Extreme Close-up} (ECU), \emph{Close-up} (CU), \emph{Medium Close-up} (MCU), \emph{Medium shot} (MS), \emph{Medium Long shot} (MLS), \emph{Long shot} (LS), \emph{Extreme Long shot} (ELS).
In practical cases, especially when an automatic computation is required, these categories are further reduced to three fundamental families \cite{BSA16}: \emph{Close shots} (CS), \emph{Medium shots} (MS), and \emph{Long shots} (LS), as shown in Figure~\ref{fig:types}.

Close shots focus on a small area of the scene, such as a face, in such a detail that it almost fills the screen.
Used to abstract the subject from a context, CS reveal important details of the plot or character's feelings.
In Medium shots actors and the setting share roughly equal space.
Eventually, Long shots show a broad view of the surroundings around the subject.

Shot scale is then described by a 12-dimensional feature vector obtained by concatenating:
\begin{itemize} 
\item the shot Scale Distribution (\emph{SDistr}) over the three classes CS, MS, and LS (3-dimensions);
\item the shot Scale Transition matrix (\emph{STrans}), which describes the probability transition in time of going from any class to another class (9-dimensions).
\end{itemize}
Shot scale feature vectors are built on a second-by-second base since an individual camera take may contain several scales whenever the camera or the objects in the image are moving.
Special cases when two different shot scales are found in the same image such as \emph{Foreground shots} (FS) and \emph{Over-the-shoulder shots} (OS) \cite{SBA15} are not considered in this analysis.

%%%%%%%%%%%%%%%%%%%%%%%%%%%%%%%%%%%%%%%%%%%%%%%%%%%%%%%%%%%%%%%%%%%%%%%%%%%%%%%%%%%%%%%%%%%%%%%%%%%%%%%%%%%%%
\section{Movie and ground-truth data}
\label{sec:database}

To understand at which level shot scale and duration are identifiers of a director's style, we analyse 120 movies directed from 1950 to 2013 by six authoritative directors: Michelangelo Antonioni, Ingmar Bergman, Federico Fellini, Jean-Luc Godard, Martin Scorsese, and Béla Tarr.
For all authors the almost complete filmography is considered, excluding only movies at low video quality.
To the best of our knowledge this is the largest, and probably unique, database for automatic analysis of art movies.
The number of representative authors is limited, but well enough to demonstrate the feasibility of authorship recognition.
The full list of movies (approximately half black-and-white (b\&w), half colors) is given in Appendix~\ref{app-A}.

For all 120 movies, ground truth (GT) values for shot scale are manually annotated, second-by-second.
The annotation procedure took months to be completed, and involved many scholars.
At the end, multiple checks have been performed by a team made up of a cinema expert and a data scientist, in order to ensure annotation consistency across different annotators.

To account for the GT on shot duration we retrieve data from \textit{Cinemetrics} \cite{cinemetrics}, an online application populated by the film scholar community collecting data about films \cite{T09}.
%Shot boundary data from Cinemetrics are available only for 77 movies, which are indicated in the last column of the table in Appendix~\ref{app-A} with their Cinemetrics ID.
%For the remaining 43 we have only the GT about shot scale.
Shot boundary data are available only for 77 movies only (those indicated in the last column of Appendix~\ref{app-A} with their Cinemetrics ID) even if there is not always a perfect temporal alignment between video files and the GT annotations.

%%%%%%%%%%%%%%%%%%%%%%%%%%%%%%%%%%%%%%%%%%%%%%%%%%%%%%%%%%%%%%%%%%%%%%%%%%%%%%%%%%%%%%%%%%%%%%%%%%%%%%%%%%%%%

\section{Author recognition based on GT features}
\label{sec:exp1}

\subsection{Visualization of authorship data}

\begin{figure*}[!ht]
     \centering
      \begin{subfigure}{0.58\columnwidth}
        \includegraphics[width=\columnwidth]{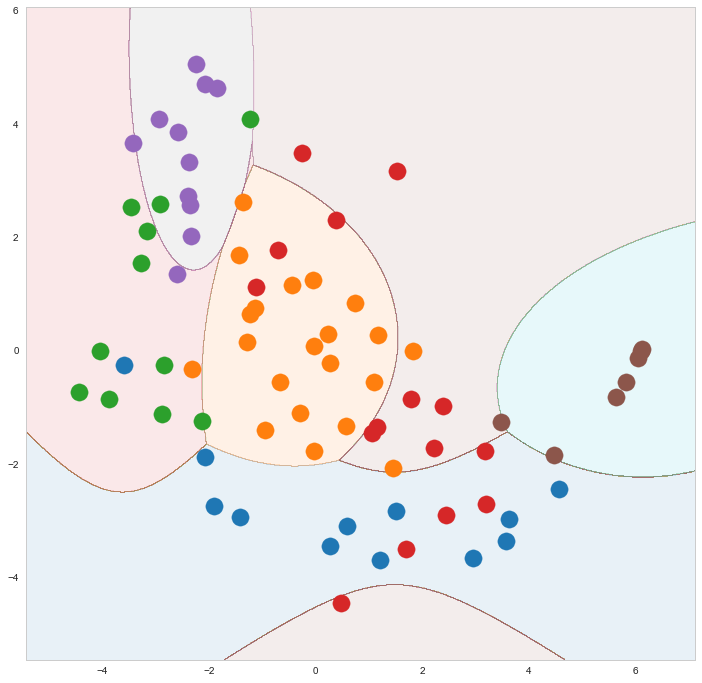}
        \caption{}
      \end{subfigure}
      %\par
      \begin{subfigure}{0.62\columnwidth}
        \includegraphics[width=\columnwidth]{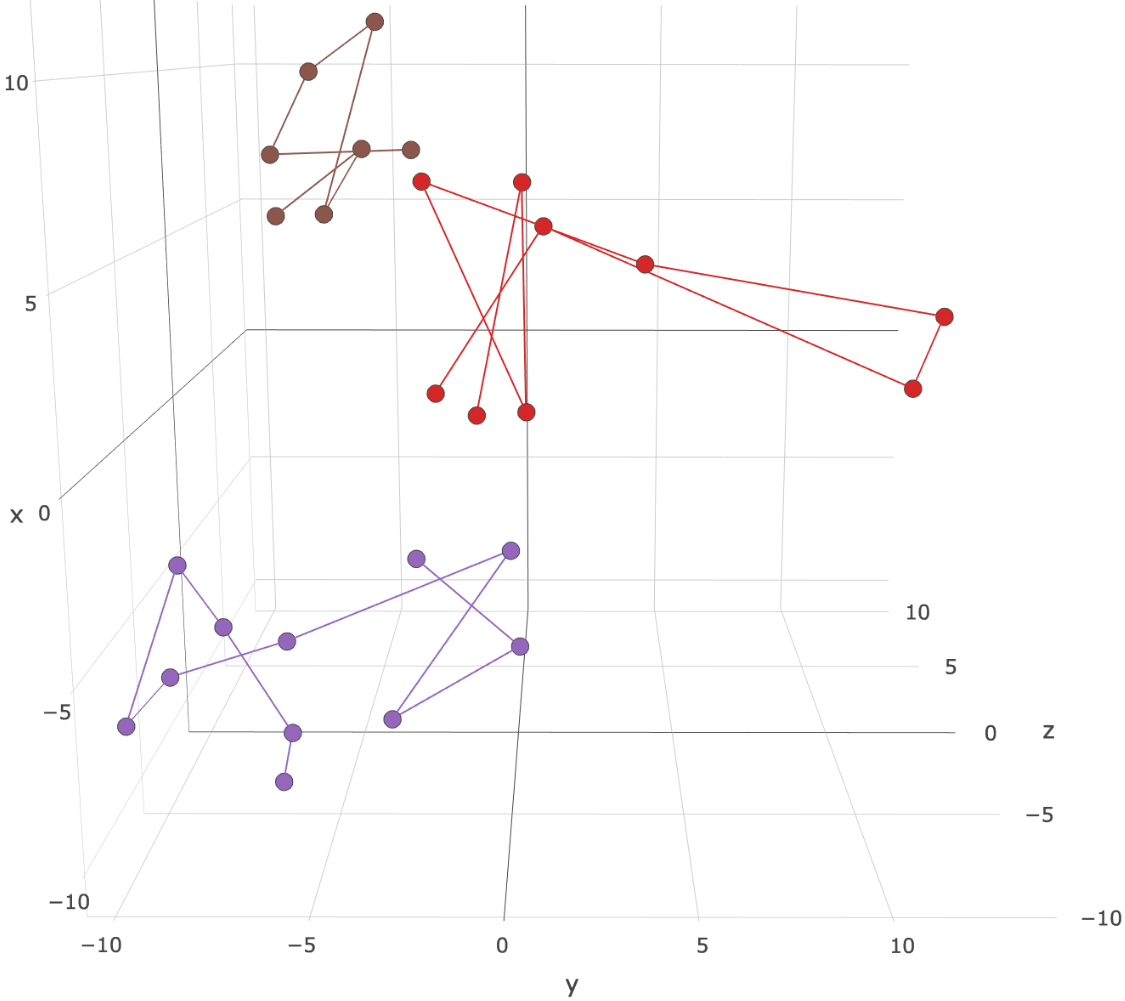}
        \caption{}
      \end{subfigure}
      %\par
      \begin{subfigure}{0.61\columnwidth}
        \includegraphics[width=\columnwidth]{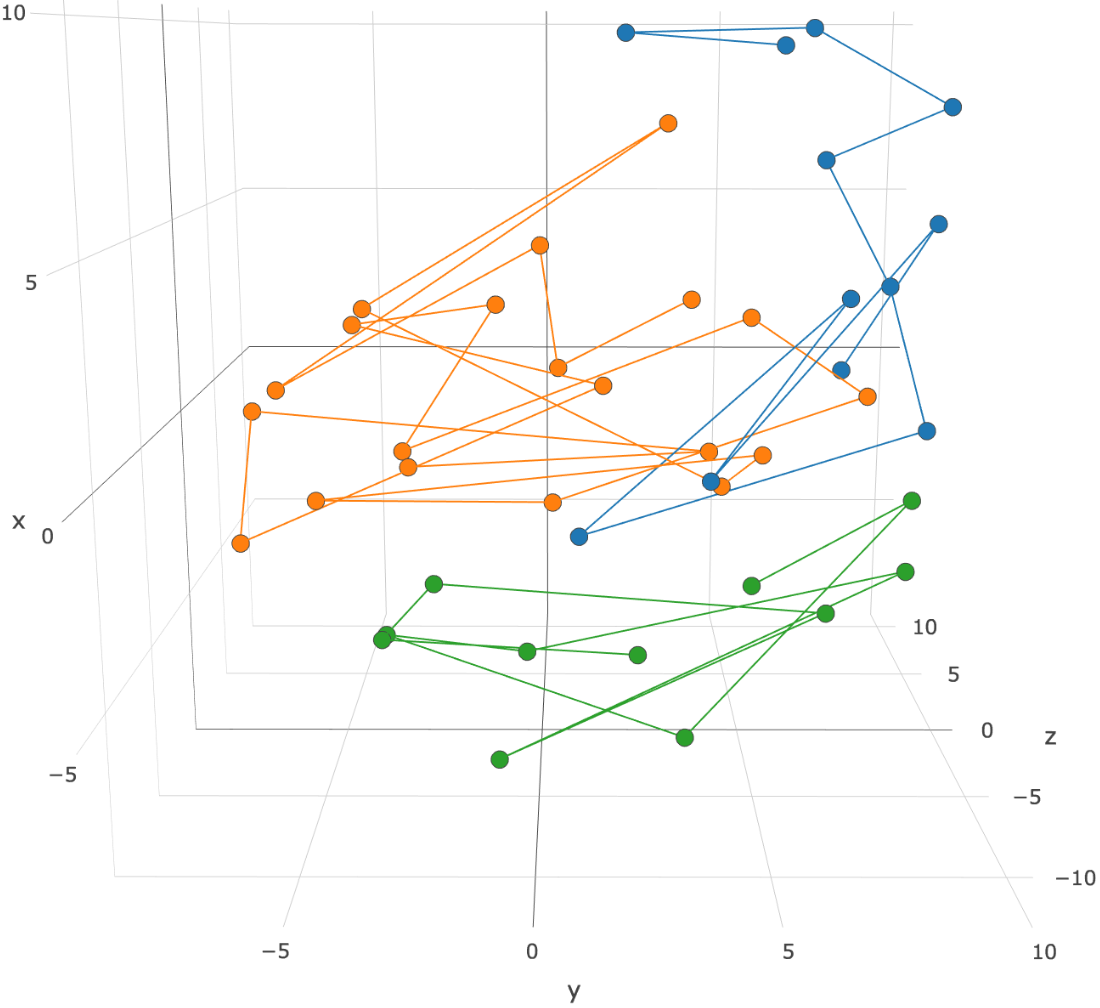}
        \caption{}
      \end{subfigure}
\caption{(a) 2D map built with t-SNE using GT shot features on 77 movies by Godard (red), Scorsese (violet), Tarr (brown), Antonioni (blue), Bergman (orange), and Fellini (green). 3D maps showing movie clusters on 3 authors, with same colours as before: (b) Godard, Scorsese, and Tarr; (c) Antonioni, Bergman, and Fellini. More interactive maps on the project page \cite{project}.}
\label{fig:t_sne_all_2D_newdata}
\end{figure*}

To preliminary assess the feasibility of an automatic procedure for authorship recognition, we visualize feature data for each author on the subset of 77 movies with complete GT.
By means of \textit{t-Distributed Stochastic Neighbor Embedding (t-SNE)} \cite{t_SNE}, a dimensionality reduction technique suited for visualisation of large datasets, we build two and three-dimensional maps in which distances between graphical points reflect data similarities.
In a feature space reduced from 68 to 2 dimensions (Figure~\ref{fig:t_sne_all_2D_newdata}(a)) we observe that movies from different authors cluster together quite well.
The ability of proposed features in separating different authors is better appreciated in the 3-dimensional t-SNE maps (Figure~\ref{fig:t_sne_all_2D_newdata}(a) and (b)) where each one represents movies of three directors from convenient view angles.
All 3D interactive graphs are available on the project website \cite{project} for further inspection.

%%%%%%%%%%%%%%%%%%%%%%%%%%%%%%%%%%%%%%%%%%%%%%%%%%%%%%%%%%%%%%%%%%%%%%%%%%%%%%%%%%%%%%%%%%%%%%%%%%%%%%%%%%%%

\subsection{Author recognition}
\label{sec:Authorship_clustering}

To aswer Q1 (\emph{are shot duration and shot scale a sufficient feature set for recognizing a director?}) we evaluate the performance of different classifiers in attributing movies to authors.
For the 77 movies with GT, we consider the shot duration features \emph{DDistr}, \emph{DTrans} and scale ones \emph{SDistr}, \emph{STrans}.

\begin{figure}[h]
\centering
	\includegraphics[width=0.65\columnwidth]{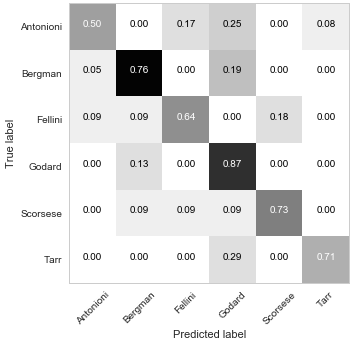}
\caption{Attribution based on GT features: confusion matrix.}
\label{fig:clust_perform_theory_CM}
\end{figure}

Employing a classical leave-one-out cross-validation, i.e. for every author all movies minus one are used for training and the leftovers for testing, we extensively test these classifiers (from \cite{scikit-learn}): Nearest Neighbors (KNN), Linear SVM, RBF-kernel SVM, Random Forest, Multi-layer Perceptron (100 hidden nodes and 3 layers), AdaBoost, and N\"aive Bayes (NB), under several coarse parameterizations.
Best three, which are linear SVM, KNN, and NB, are then accurately tuned by means of a grid search procedure with cross-validation.
The fact that SVM, and even more KNN and NB, perform better than other classifiers, is probably due to the relative scarcity of training data.
Overall best results are obtained with the Gaussian N\"aive Bayes algorithm \cite{Z04} and are presented for every author in Table~\ref{tab:clust_perform_theory} and by the confusion matrix in Figure~\ref{fig:clust_perform_theory_CM}.

\begin{table}[ht]
\centering
      \caption{Authorship recognition based on GT features (N\"aive Bayes, leave-one-out cross-validation).}
 \begin{tabular}{ l | c |  c | c |  c  }
    \hline
    \textbf{author} & \textbf{precision} &   \textbf{recall}  & \textbf{f1-score} &   \textbf{movies}\\ \hline \hline
    Antonioni    &   0.75    &  0.50   &   0.60    &    12 \\ \hline
    Bergman     &  0.80   &   0.76   &   0.78    &   21\\ \hline
    Fellini    &   0.70   &   0.64  &    0.67  &      11\\ \hline
    Godard    &   0.57   &   0.87   &   0.68   &    15 \\ \hline
     Scorsese   &    0.80   &   0.73   &   0.76   &    11\\ \hline
      Tarr   &    0.83    &  0.71  &    0.77  &     7\\ \hline \hline
\textbf{avg / total}  &     \textbf{0.74}   &   \textbf{0.71}   &   \textbf{0.71}  &    \textbf{77}\\ \hline
  \end{tabular}
    \label{tab:clust_perform_theory}
  \end{table}

Apart from Fellini and Antonioni which show slightly lower recall, all authors are recognized with high accuracy and good balance between recall and precision.
Obtained results, proving the separability properties of the proposed shot feature set, are a strong indicator of the feasibility  of automatic recognition of movie authorship by means of automatically computed shot features.
Interestingly, by using shot scale features annotated on seven classes as in \cite{Ko14} (i.e. ECU, CU, MCU, MS, MLS, LS, ELS) all scores increase by approximately $5\%$: precision=$0.79$, recall=$0.77$ and f1=$0.76$.

%%%%%%%%%%%%%%%%%%%%%%%%%%%%%%%%%%%%%%%%%%%%%%%%%%%%%%%%%%%%%%%%%%%%%%%%%%%%%%%%%%%%%%%%%%%%%%%%%%%%%%%%%%%%%

\section{Insights on shot features}
\label{sec:insights}

We carry out further analysis on single features to better understand their relevance in classification, feature correlation, dependency with respect to the year of production, and the most distinctive feature for each director (Q2).

%%%%%%%%%%%%%%%
\subsection{Single feature analysis}
\label{sec:single_feature_analysis_theory}

%%%

\begin{figure*}[!ht]
     \centering
      \begin{subfigure}{1.3\columnwidth}
        \includegraphics[width=\columnwidth]{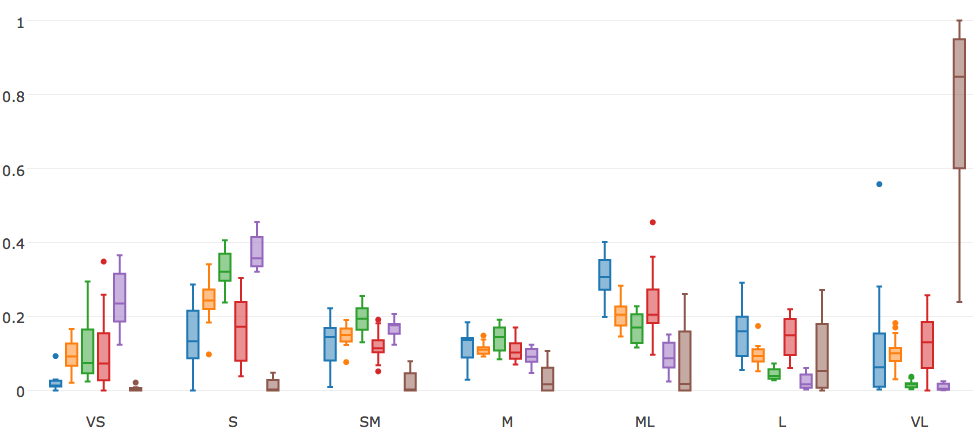}
        \caption{}
      \end{subfigure}
      \begin{subfigure}{0.7\columnwidth}
      \vspace{0.0cm}
        \includegraphics[width=\columnwidth, height=5.05cm]{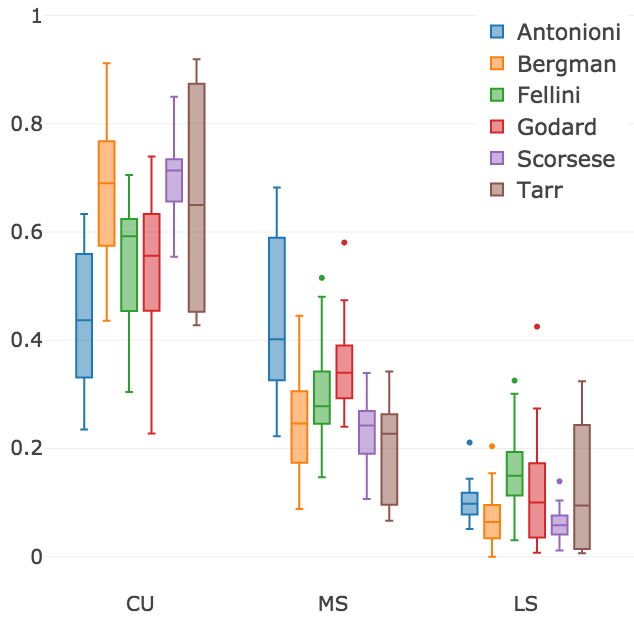}
        \caption{}
      \end{subfigure}
\caption{(a) Shot duration distributions (\emph{DDistr}) for each director on a total of 77 movies (with median, std, variance, and outliers). (b) Shot scale distributions (\emph{SDistr}) for each director on 120 movies (with median, std, variance, and outliers).}
\label{fig:SDD+SSD_all_movies}
\end{figure*}

\paragraph{Shot Duration Distribution (DDistr)}
the distributions of shot duration are shown in Figure~\ref{fig:SDD+SSD_all_movies}(a) for each director.
%
%\begin{figure*}[!ht]
%\centering
%	\includegraphics[width=2\columnwidth]{SDD_all_movies-2}\label{fig:SDD_all_movies_withTarr}
%\caption{Shot duration distributions (\emph{DDistr}) for each director on 77 movies (with median, std, variance, and outliers).}
%\label{fig:SDD_all_movies_withTarr}
%\end{figure*}
%
%
Béla Tarr's preference for VL shots is evident, as expected from cinema literature \cite{K13}.
Antonioni and Godard share a tendency to use more ML, L, and VL with respect to Fellini and Bergman, who also present similar distributions.
Conversely, Scorsese reveals little use of long shots, especially in his more recent production, when movies have become on average ``\emph{quicker, faster, and darker}'' \cite{CBD11}. 
To measure its supposed discrimination capability, by using \emph{DDistr} alone, classification accuracy (with Na\"ive Bayes classifier and leave-one-out cross-validation on the 77 movie set) scores a significant $\sim70\%$.

%%%

\begin{figure}[h]
	\includegraphics[width=\columnwidth]{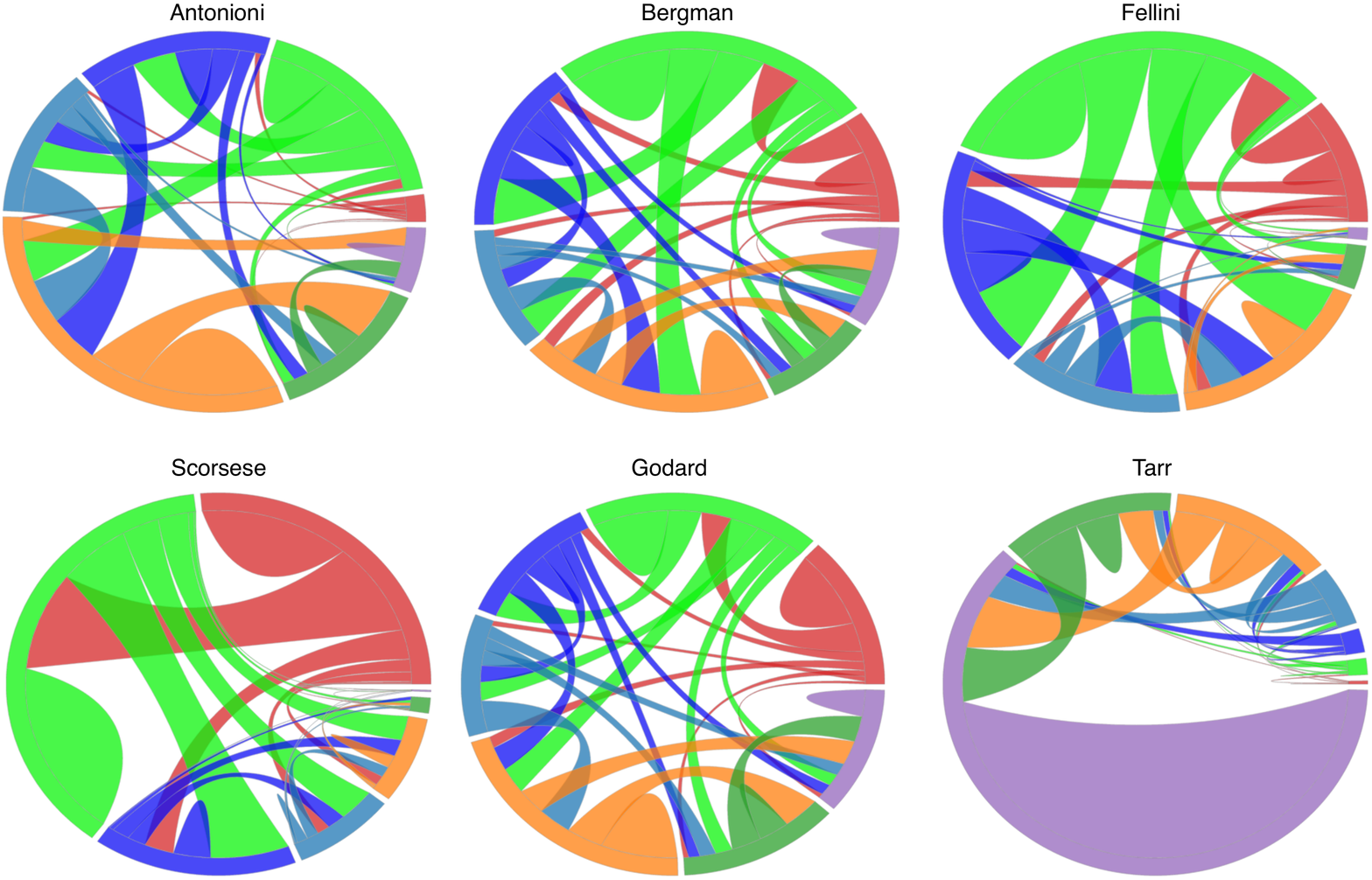}
\caption{\emph{DTrans} obtained on 77 movies with GT features. In the outer ring the distribution among shot duration classes: VS (red), S (light green), SM (blue), M (light blue), ML (orange), L (dark green), and VL (violet). Transitions between classes are represented by the connecting arcs, where the chord widths at the outer ring represent transition probabilities.}
\label{fig:SDT_union}
\end{figure}

\paragraph{Shot Duration Transition (DTrans)}
Figure~\ref{fig:SDT_union} represents the information related to transitions between shot durations for each director on 77 movies by means of chord diagrams.
From an observation it is evident that the probability to move to a shot with different duration is quite high, but often on contiguous classes, meaning that the scene slightly speed ups or decelerates.
As a consequence, connecting arcs from long to short duration shots (and viceversa) are seldom registered.
Since diagrams look very different from author to author, we suppose that this is a highly distinctive author feature.
In fact by using \emph{DTrans} as a solo feature, classification (NB classifier on 77 movies) reaches $\sim63\%$ in accuracy.

%%%
\paragraph{Shot Scale Distribution (SDistr)}
the distributions of shot scale for each director on all 120 movies are shown in Figure~\ref{fig:SDD+SSD_all_movies}(b), where it is evident the dominance of Close shots on other scales.
%
%\begin{figure}[h]
%\centering
%	\includegraphics[width=\columnwidth]{SSD_all_movies-2.png}
%\caption{Shot scale distributions (\emph{SDistr}) for each director on 120 movies (with median, std, variance, and outliers).}
%\label{fig:SSD_all_movies}
%\end{figure}
%
%
%
Despite small individual differences, such as Antonioni's preference for MS or Scorsese's low standard deviation on scale usage, no director's specific patterns are clearly visible.
Confirming this impression, by using only the \emph{SDistr} feature, and adopting the procedure of the previous experiments, classification accuracy drops to $\sim27\%$.
However, by using the more refined shot scale categorization on seven scales as in \cite{Ko14} performance increases up to $\sim48\%$.

%%%

\begin{figure}[h]
\centering
	\includegraphics[width=\columnwidth]{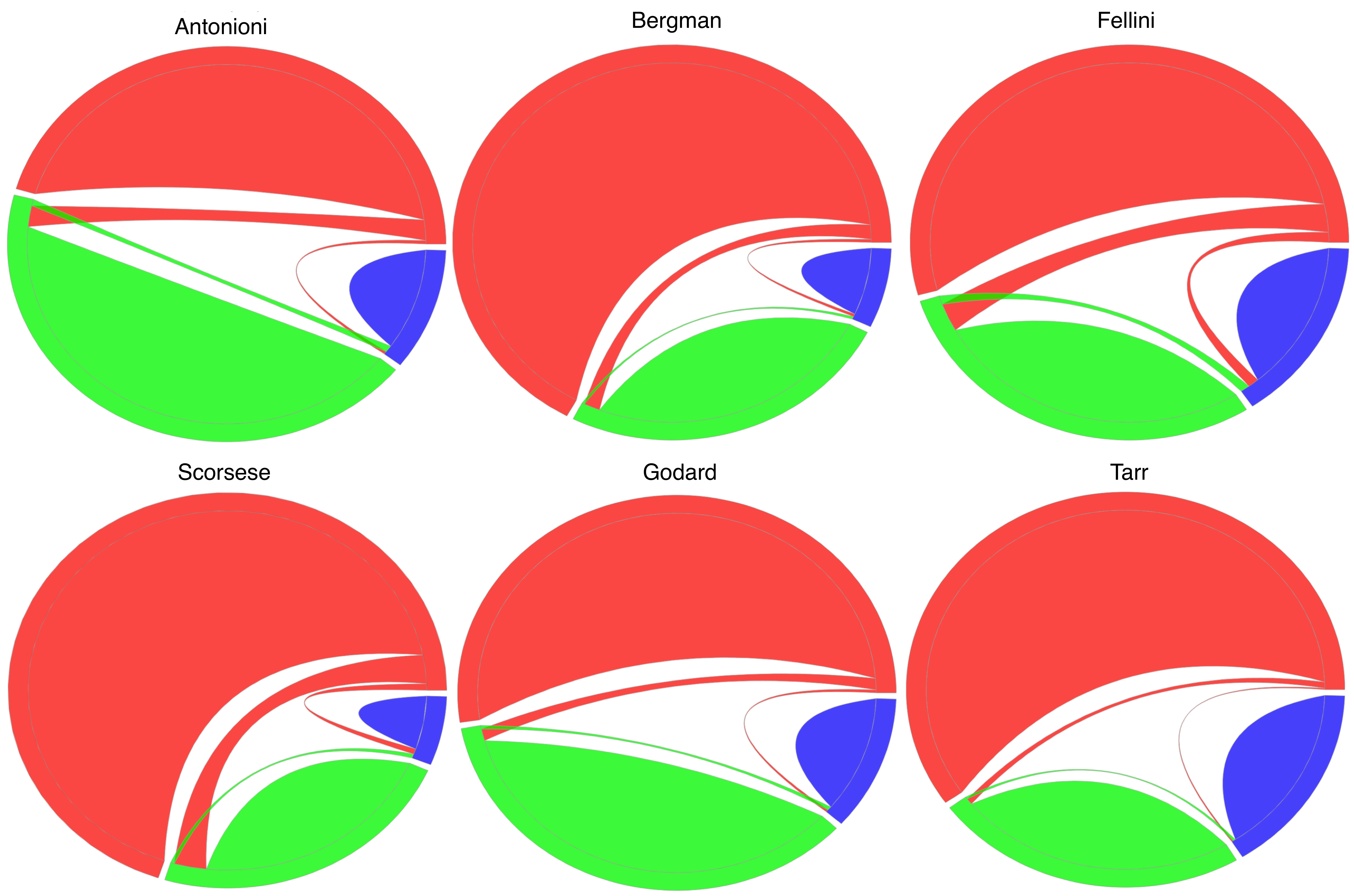}
\caption{\emph{STrans} for each director on 120 movies based on GT features. In the outer ring the distribution among scale classes: CS (red), MS (green), and LS (blue). Transitions between classes are represented by the connecting arcs, where the chord widths at the outer ring represent transition probabilities.}
\label{fig:SST_union}
\end{figure}

\paragraph{Shot Scale Transition (STrans)}
Figure~\ref{fig:SST_union} represents the information related to transitions between shot scales for each director by means of chord diagrams.
Scale data are arranged radially around a circle (CS in red, MS in green, LS in blue) and the transition probabilities between scales are represented as connecting arcs.
Since scale is annotated each second, transition probabilities to different scales are pretty small. 
However \emph{STrans} feature is quite informative, so that using it as \emph{a solo} feature, the accuracy jumps to $\sim60\%$, doubling the performance obtained using \emph{SDistr}.

%%%

To wrap up the analysis on single features and their classification abilities when taken alone, we report in Table~\ref{tab:combined-feature-performance} the accuracy scores obtained by using them as \emph{a solo} features, where it is evident the superior ability in classification by duration features with respect to scale related ones.

\begin{table}[ht]
\centering
      \caption{Authorship classification obtained by single features. Performance for \emph{SDistr} and \emph{STrans} are given considering a shot scale subdivision in 3 and 7 different classes.}
 \begin{tabular}{ l | c   }
    \hline
    \textbf{feature} & \textbf{accuracy }  \\ \hline  \hline
\textit{DDistr} & 0.701\\ 
\textit{DTrans} & 0.636 \\ 
\hline
\hline
\textit{SDistr} (3-scales) & 0.272 \\ 
\textit{STrans} (3-scales) & 0.597 \\ 
\hline
\textit{SDistr} (7-scales) & 0.480 \\ 
\textit{STrans} (7-scales) & 0.584 \\ 
\hline
  \end{tabular}
    \label{tab:combined-feature-performance}
  \end{table}

\begin{figure}[h]
\centering
	\includegraphics[width=\columnwidth]{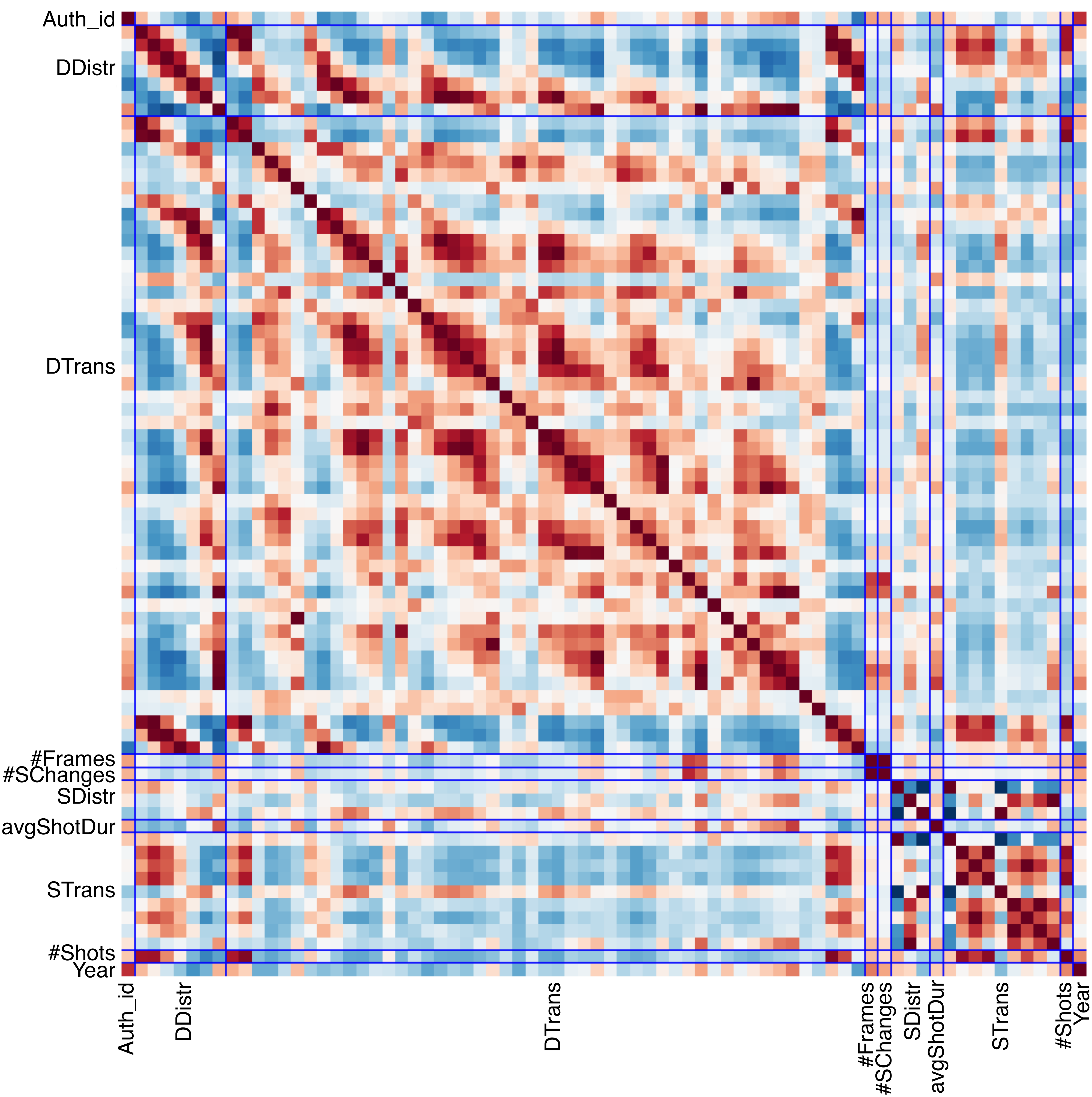}
\caption{Correlation matrix between all feature components.}
\label{fig:feat-corr}
\end{figure}

Figure~\ref{fig:feat-corr} depicts the correlation matrix between all feature components (arranged in blocks).
To investigate other possible relations between features and authors, correlations have been computed also for other \emph{accessory features} we could easily obtain, such as the \emph{year} of production, and other easily derived features such as the total number of movie shots (\#\textit{Shots}), the movie total duration (\#\textit{Frames}), and the number of scale changes in the movie (\#\textit{SChanges}), which we  investigated in \cite{CBL11} as a responsible for triggering audience's emotional involvement.
From the inspections of Figure~\ref{fig:feat-corr} no other conspicuous correlations between feature blocks emerge.

We have also tried to exploit these accessory features in classification, obtaining no significant difference with respect to results in Table~\ref{tab:clust_perform_theory}, that is a $+1\%$ in accuracy on 3 scales, and no increase at all on 7 scales.
By adding the \emph{year} of production to our feature set, we obtain a slight increase in accuracy of about $2\%$.
Other relations between the feature set and the year of production are investigated further on.

\subsection{Feature temporal analysis}
\label{sec:temporal_characterisation}

\begin{figure}[h]
\centering
	\includegraphics[width=0.95\columnwidth]{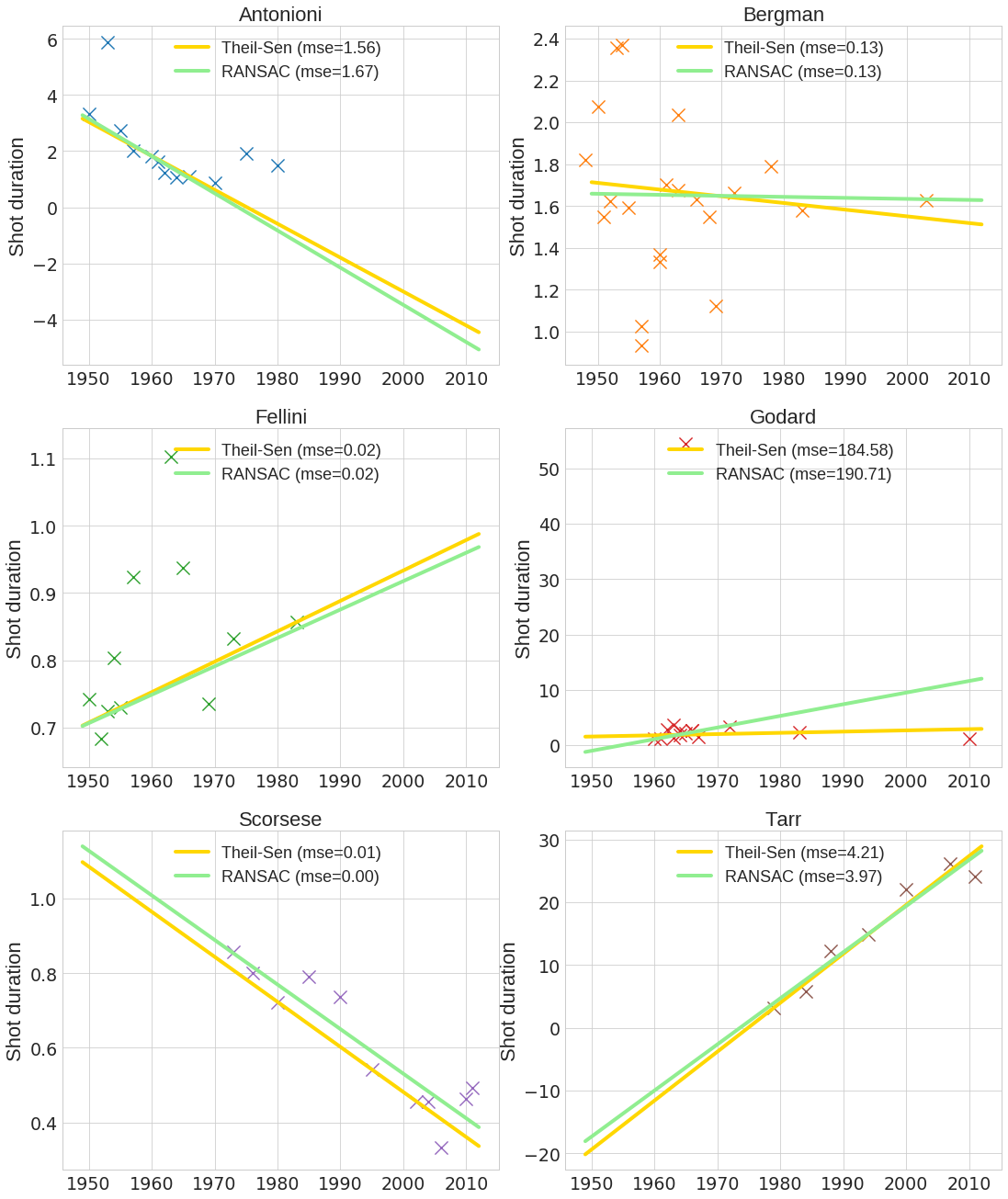}
\caption{Trends of shot duration for each director (obtained by Ransac \cite{FB81} and Theil-Sen \cite{O15} for their resilience to outliers).}
\label{fig:average_shot_duration_VS_year_newdata+std}
\end{figure}

Movies have changed dramatically over the last century, with several reflections on film style, especially on the increased shot pace, on motion, and luminance \cite{C16a}.
However, if we restrict the analysis to the shot duration of our GT data, there is no much evidence of such trend (see Figure~\ref{fig:average_shot_duration_VS_year_newdata+std}).

Scorsese is the only one who significantly decreases his mean shot duration in years.
It is also true that he is the only one (with Tarr, who is obviously an outlier with respect to duration) who lately directed a consistent number of movies (production years are in Appendix~\ref{app-A}).
%
%\begin{figure}[h]
%\centering
%	\includegraphics[width=\columnwidth]{gantt-directors-3}
%\caption{Years and productivity of the considered directors.}
%\label{fig:gantt-directors}
%\end{figure}
%
Even if a thorough temporal analysis of all features transcends the objectives of this work, since there is a natural high correlation between the year of production and the author, it is important to exclude that the proposed feature set is related to the production period instead of being specific of the director.
In order to analyse this aspect, we automatically infer the historical period of movie production.
By adopting the same procedure with N\"aive Bayes classifier and leave-one-out cross-validation on the 77 GT movies, we obtain results presented in Table~\ref{tab:year-estimation}.
These exclude that the proposed features are related to the period (we have considered decades of production), meaning that the classifier does not learn the author from the year.

\begin{table}[ht]
\centering
      \caption{Prediction of the production period based on GT.}
 \begin{tabular}{ c | c |  c | c |  c  }
    \hline
    \textbf{period} & \textbf{precision} &   \textbf{recall}  & \textbf{f1-score} & \textbf{movies}  \\ \hline \hline
    (0) 1948-1959    &   0.29    &  0.21  &    0.24    &   20\\ \hline
    (1) 1960-1969     &   0.43   &   0.43   &   0.43   &   28 \\ \hline
    (2) 1970-1979    &    0.06  &    0.11   &   0.08    &   9 \\ \hline
    (3) 1980-1989    &  0.00   &   0.00  &    0.00  &     7\\ \hline
    (4) 1990-1999  &    0.00   &   0.00   &   0.00   &    3 \\ \hline
    (5) 2000-2009   &    0.40   &   0.33   &   0.36  &      6 \\ \hline 
    (6) 2010-2017   &    0.14  &    0.25   &   0.18   &    4  \\ \hline  \hline
	\textbf{avg / total}  &     \textbf{0.27}   &   \textbf{0.26}   &   \textbf{0.26} &   \textbf{77} \\ \hline
  \end{tabular}
    \label{tab:year-estimation}
  \end{table}

\subsection{Relevance of features in different authors}
\label{ssec:feature_for_director}

\begin{figure}[!ht]
     \centering
      \begin{subfigure}{0.9\columnwidth}
        \includegraphics[width=\columnwidth]{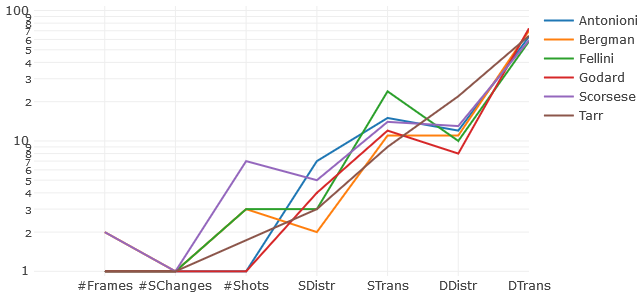}
        \caption{in absolute value}
      \end{subfigure}
      %\par
      \begin{subfigure}{0.9\columnwidth}
        \includegraphics[width=\columnwidth]{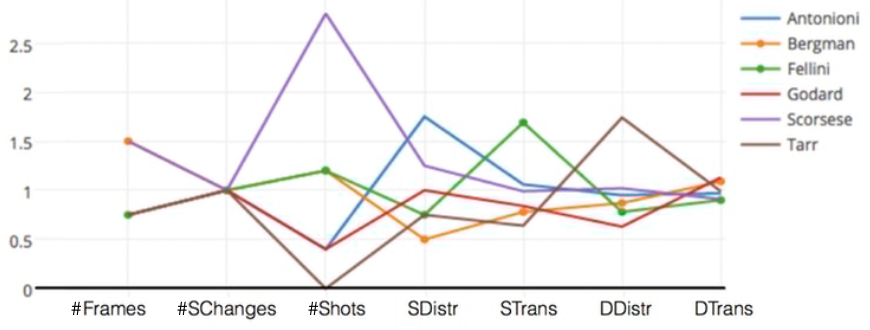}
        \caption{normalized to the mean across authors}
      \end{subfigure}
      %\par
\caption{Importance measure of single features for each author.}
\label{fig:feature_weights}
\end{figure}

To verify if some features are more relevant than others for recognizing an author, we train a classifier for each director with the one-against-all approach (i.e. putting movies from one director in the positive class and those from other authors in the negative one).
For assessing how much a certain feature is specific of an author, we exploit the \textit{feature importance measure} returned by the Random Forest implementation in \cite{scikit-learn}, which evaluates how much each feature decreases the weighted impurity in a tree.
The importance of \emph{SDistr}, \emph{STrans}, \emph{DDistr}, \emph{DTrans}, and accesory \#\textit{Frames}, \#\textit{Shots}, and \#\textit{SChanges} are given in Figure~\ref{fig:feature_weights}(a) in absolute values, and in Figure~\ref{fig:feature_weights}(b) normalized to their mean value across directors.

Figure~\ref{fig:feature_weights}(a) confirms that shot duration features are in general more important than those related to scale.
For all directors, transition features (\emph{STrans} and \emph{DTrans}) have higher impact than the related distributions (\emph{SDistr} and \emph{DDistr}).
Accessory features (\#\textit{Frames}, \#\textit{Shots}, and \#\textit{SChanges}) are, in absolute terms, of lower importance.

In Figure~\ref{fig:feature_weights}(b) feature importance measures are normalized to the mean across authors, thus highlighting individual differences.
Scorsese differs from others for the total number of movie shots.
Scale distribution (\emph{SDistr}) is the most specific feature for Antonioni instead, as expected from the study in \cite{Ko14}, while Fellini most distinguishing feature is the transition pattern between shot scales (\emph{STrans}).
Bergman and Godard do not present any peculiar feature, and this is probably why their authorships are sometimes confused in classification, as shown in Figure~\ref{fig:clust_perform_theory_CM}.
As expected, Tarr is characterised by his unique use of Long shots, which affects his \emph{DDistr}.

%%%%%%%%%%%%%%%%%%%%%%%%%%%%%%%%%%%%%%%%%%%%%%%%%%%%%%%%%%%%%%%%%%%%%%%%%%%%%%%%%%%%%%%%%%%%%%%%%%%%%%%%%%%%%
\section{Automatic shot feature analysis}
\label{sec:automatic_shot_analysis}

We now perform the task of author recognition by means of automatic procedures.
Shot transitions are detected by existing methods, while we propose a novel deep learning approach to derive shot scale.
Eventually we compare performance on author recognition with results obtained using GT features.

%%%%%%%%%%%%%%%%%%%%%%%%%%%%%%%%%%%%%%%%%%%%%%%%%%%%%%
\subsection{Shot duration features}

For temporal segmentation of videos into shots we compare the algorithm in \cite{AM14} and a recent implementation of the one in \cite{AL99} extended to identify cuts.
The first method detects abrupt and gradual transitions based on frame similarity computed by means of both local (SURF) and global (HSV histograms) descriptors, while the second one exploits histogram information and decision criteria derived from statistical properties of cuts, dissolves, and wipes.
To select the algorithm, we test both methods on \emph{Red desert (Antonioni, 1964)}, which counts more than 700 transitions, obtaining the performance in Table~\ref{tab:shotcut}.

\begin{table}[ht]
\centering
      \caption{Comparison of shot boundary detectors.}
 \begin{tabular}{ l | c |  c | c   }
    \hline
    \textbf{method} & \textbf{precision} &   \textbf{recall}  & \textbf{f1-score}   \\ \hline \hline
     Apostolidis et. al. \cite{AM14}  &   \textbf{93.58}    &  91.00  &    92.27    \\ \hline
     Adami et. al. \cite{AL99}     &    89.16  &    \textbf{97.01}   &   \textbf{92.91}     \\ \hline  
  \end{tabular}
    \label{tab:shotcut}
  \end{table}

\begin{figure}[!ht]
     \centering
      \begin{subfigure}{0.49\columnwidth}
        \includegraphics[width=\columnwidth]{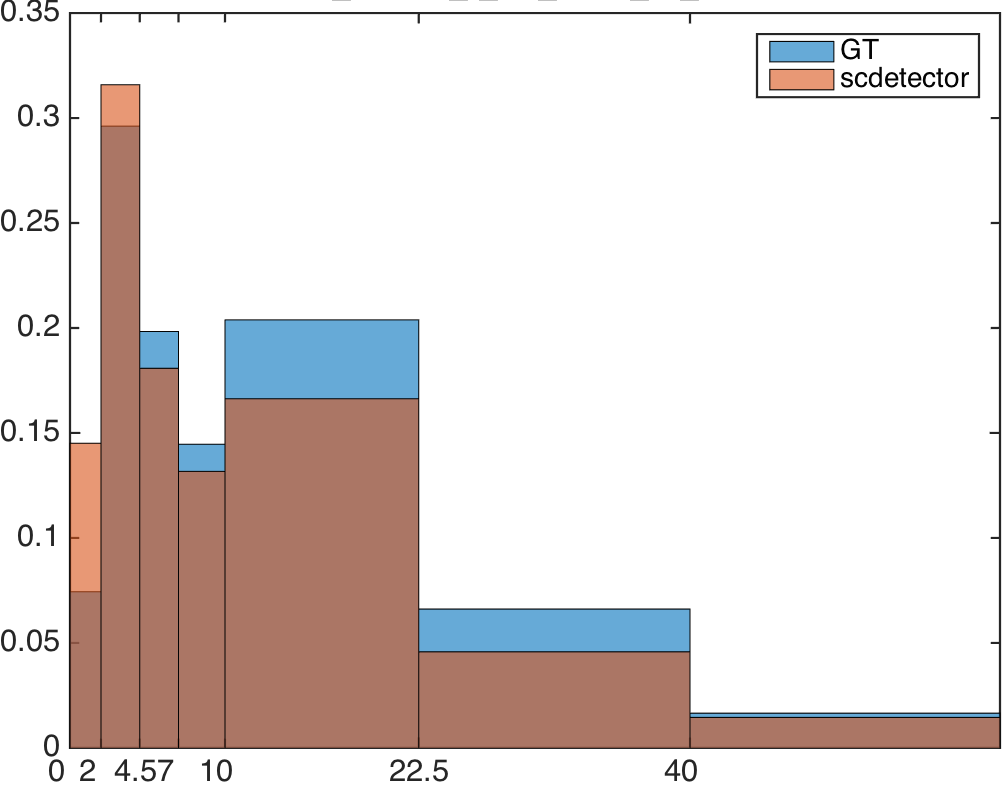}
        \caption{\textit{Le notti di Cabiria}}
      \end{subfigure}
      %\par
      \begin{subfigure}{0.49\columnwidth}
        \includegraphics[width=\columnwidth]{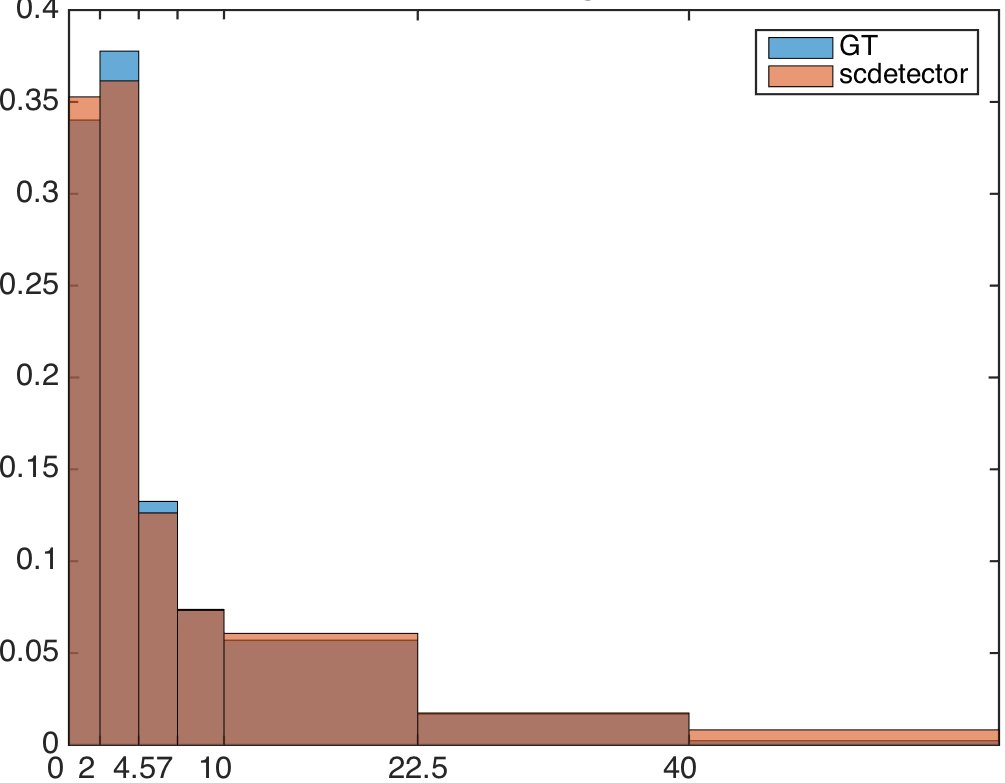}
        \caption{\textit{Gangs of New York}}
      \end{subfigure}
      %\par
\caption{Automatically computed shot duration distributions (in red) and GT ones (in blue) for (a) \textit{Le notti di Cabiria (Fellini, 1957)}, and (b) \textit{Gangs of New York (Scorsese, 2002)}.}
\label{fig:DDistr_comparison}
\end{figure}

Scores are computed with respect to a manually annotated GT at frame level, confirming the efficiency of both methods.
Since the method in \cite{AM14} is pretty slow (code here \cite{shotcut-code}), while the recent implementation of \cite{AL99}, named \emph{scdetector}, works in real-time, we use the latter for producing \emph{DDistr} and \emph{DTrans} features on the whole set of 120 movies.
%As examples of automatically computed features, we show in Figure~\ref{fig:DDistr_comparison} the duration distributions obtained by \cite{AL99} of \textit{Le notti di Cabiria (Fellini, 1957)}  and \textit{Gangs of New York (Scorsese, 2002)}, overlaid to GT ones already shown in Figure~\ref{fig:SDD_example}.
We show in Figure~\ref{fig:DDistr_comparison} the duration distributions obtained by \cite{AL99} of \textit{Le notti di Cabiria (Fellini, 1957)} and \textit{Gangs of New York (Scorsese, 2002)} (in red), overlaid to GT ones (in blue).

For most movies, automatically derived \emph{DDistr} show little deviation from the corresponding GT distributions, with few problems only on Godard's production.
We show these results aggregated for each author in Table~\ref{tab:DDT-hist-comparison}, where we compare computed features with GT ones only in terms of \emph{correlation} and \emph{histogram intersection} on all 77 GT movies (both normalized in [0,1]).
No accuracy based measures are possible in this case, since there is often a temporal misalignment between our video files (on which we run the shot boundary method) and the GT annotations from Cinemetrics.

\begin{table}[ht]
\centering
      \caption{Automatically computed \emph{DDistr} compared with GT ones in terms of correlation and histogram intersection.}
 \begin{tabular}{ l | c |  c |  c  }
    \hline
    \textbf{author} & \textbf{correlation (std)} &   \textbf{hist-intersection (std)}   &   \textbf{movies}\\ \hline \hline
    Antonioni    &   0.946$\pm$0.096    &  0.907$\pm$0.095     &   12 \\ \hline
    Bergman     &  0.832$\pm$0.191   &   0.911$\pm$0.050     &  21 \\ \hline
    Fellini    &   0.934$\pm$0.088   &   0.909$\pm$0.056   &     11 \\ \hline
    Godard    &   0.593$\pm$0.507   &   0.836$\pm$0.119      &    15\\ \hline
     Scorsese   &    0.986$\pm$0.014   &   0.956$\pm$0.030   &    11\\ \hline
      Tarr   &    0.948$\pm$0.098    &  0.771$\pm$0.094  &  7   \\ \hline 
        \end{tabular}
    \label{tab:DDT-hist-comparison}
  \end{table}

Once computed shot boundaries, \emph{DTrans} are easily derived.
Chord diagrams representing duration transitions are not shown here for the sake of brevity, but show no substantial difference with respect to those in Figure~\ref{fig:SDT_union}.

%%%%%%%%%%%%%%%%%%%%%%%%%%%%%%%%%%%%%%%%%%%%%%%%%%%%%%
\subsection{Shot scale features}

In \cite{BSA16} we first propose an automatic framework for estimating the scale by using inherent characteristics of shots such as \emph{colour intensity}, \emph{motion}, \emph{perspective}, \emph{human presence} and \emph{spectral representation}.
Experiments conducted on Antonioni's movies confirmed the validity of the framework.

In this work we propose a novel approach for shot scale estimation on the full filmographies of the six directors based on deep learning techniques.
To this aim we select three well known networks with increasing capacity: AlexNet \cite{KSH12}, GoogleLeNet \cite{SLJ15}, and VGG-16 \cite{SZ14a}.
We train the three DL architectures by using 55 movies of the dataset.
Test movies are the other 65 (marked with $^*$ in Appendix~\ref{app-A}) chosen, whenever possible, by uniformly sampling director's productions, and trying to balance b\&w and colour movies.

\begin{figure}[!ht]
     \centering
      \begin{subfigure}{0.315\columnwidth}
        \includegraphics[width=\columnwidth]{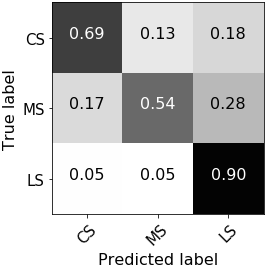}
        \caption{AlexNet}
      \end{subfigure}
      %\par
      \begin{subfigure}{0.315\columnwidth}
        \includegraphics[width=\columnwidth]{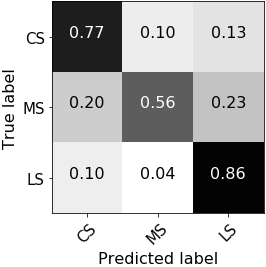}
        \caption{GoogleLeNet}
      \end{subfigure}
      %\par
      \begin{subfigure}{0.315\columnwidth}
        \includegraphics[width=\columnwidth]{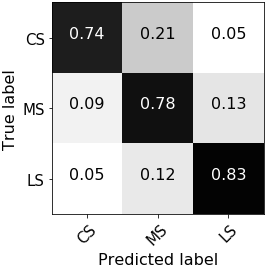}
        \caption{VGG-16}
      \end{subfigure}
      %\par
\caption{Comparison on shot scale classification (65 movies): (a) AlexNet \cite{KSH12}; (b)  GoogleLeNet \cite{SLJ15}; (c) VGG-16 \cite{SZ14a}.}
\label{fig:SDist-comparison-DL}
\end{figure}

For each Convolutional Neural Network (CNN) we adopt four different configurations: train from scratch, load weights from ImageNet \cite{imagenet} and finetune the last layer only, finetune all fully-connected layers, and finetune the whole network.
All CNNs perform better using precomputed weights from ImageNet, which is somehow expected since frame visual content is not much dissimilar from the one in ImageNet.
The three architectures have almost same performance when only the last layer, or all the fully-connected ones are trained, while a gain of $+1.5$ point is obtained on average by finetuning the whole net.
Two different networks (per each architecture) are trained to deal with color and b\&w movies, respectively.
The overall best model is VGG-16 trained with stochastic gradient descent with momentum 0.9 and weight decay 5e-7. 
Learning is carried on for 30 epochs with a base learning rate of 1e-5, divided by 10 in case of validation error plateaus.
To exploit the temporal correlation of the shot scale, results obtained each second are averaged by a moving window of 3 s.
Table~\ref{tab:DL-perf} and Figure~\ref{fig:SDist-comparison-DL} show the scale classification results and the related confusion matrices, highlighting the superiority of VGG-16.

\begin{table}[ht]
\centering
      \caption{Shot scale classification with three DL networks.}
 \begin{tabular}{ c | c |  c | c |  c  }
    \hline
    \textbf{AlexNet \cite{KSH12}} & \textbf{precision} &   \textbf{recall}  & \textbf{f1-score} &   \textbf{frames}\\ \hline
    CS  &     0.87  &    0.69    &  0.77  &  228909 \\ \hline
    MS   &    0.66   &   0.54   &   0.60   & 117570 \\ \hline
    LS     &  0.46   &   \textbf{0.90}   &   0.61 &    71644 \\ \hline
   avg / total    &   0.74   &   0.68  &    0.69  &  418123\\    \hline \hline
   \textbf{GoogleLeNet \cite{SLJ15}} &  &    & &   \\ \hline
    CS    &   0.86  &    \textbf{0.77}   &   0.81  &  238757 \\ \hline
    MS    &   \textbf{0.71}   &   0.56   &   0.63  &  116490 \\ \hline
     LS    &  0.48   &   0.86   &   0.62   &  62876 \\ \hline
       avg / total    &   0.76  &   0.73  &    0.73  &  418123\\ \hline \hline
\textbf{VGG-16 \cite{SZ14a}} &  &     &  &  \\ \hline
              CS  &     \textbf{0.92}   &   0.74   &   \textbf{0.82}  &  232547 \\ \hline
    MS   &   0.66  &    \textbf{0.78}    &  \textbf{0.72}  &  135001 \\ \hline
     LS   &   \textbf{0.60}   &   0.83   &   \textbf{0.70}   &  50575 \\ \hline
   avg / total    &   \textbf{0.80}  &   \textbf{0.77}  &    \textbf{0.77}  &  418123\\ \hline 
     \end{tabular}
    \label{tab:DL-perf}
  \end{table}

For most movies, automatically derived scale distributions by VGG-16 show little deviation from the corresponding GT distributions.
We show this result aggregated for each author in Table~\ref{tab:SDistr-hist-comparison}, where we compare automatically computed \emph{SDistr} features with GT ones in terms of \emph{correlation} and \emph{histogram intersection} metrics on all 65 tested movies.
\begin{table}[ht]
\centering
      \caption{Automatically computed \emph{SDistr} compared with GT ones in terms of correlation and histogram intersection.}
 \begin{tabular}{ l | c |  c |  c  }
    \hline
    \textbf{author} & \textbf{correlation (std)} &   \textbf{hist-intersection (std)}   &   \textbf{movies}\\ \hline \hline
    Antonioni    &   0.849$\pm$0.362    &  0.849$\pm$0.105     &   9 \\ \hline
    Bergman     &  0.950$\pm$0.068   &   0.777$\pm$0.113       &  21 \\ \hline
    Fellini    &   0.731$\pm$0.325   &   0.645$\pm$0.129    &     7 \\ \hline
    Godard    &   0.661$\pm$0.448   &   0.642$\pm$0.251      &    12\\ \hline
     Scorsese   &    0.919$\pm$0.155   &   0.649$\pm$0.122     &    11\\ \hline
      Tarr   &    0.962$\pm$0.045    &  0.952$\pm$0.050   &  5   \\ \hline 
        \end{tabular}
    \label{tab:SDistr-hist-comparison}
  \end{table}
Once computed the shot scale for each movie, \emph{STrans} features are easily derived.
Chord diagrams representing scale transitions are not shown here for the sake of brevity, but show no substantial difference with respect to those in Figure~\ref{fig:SST_union}.

\begin{figure}[!ht]
     \centering
      \begin{subfigure}{0.50\columnwidth}
        \includegraphics[width=\columnwidth]{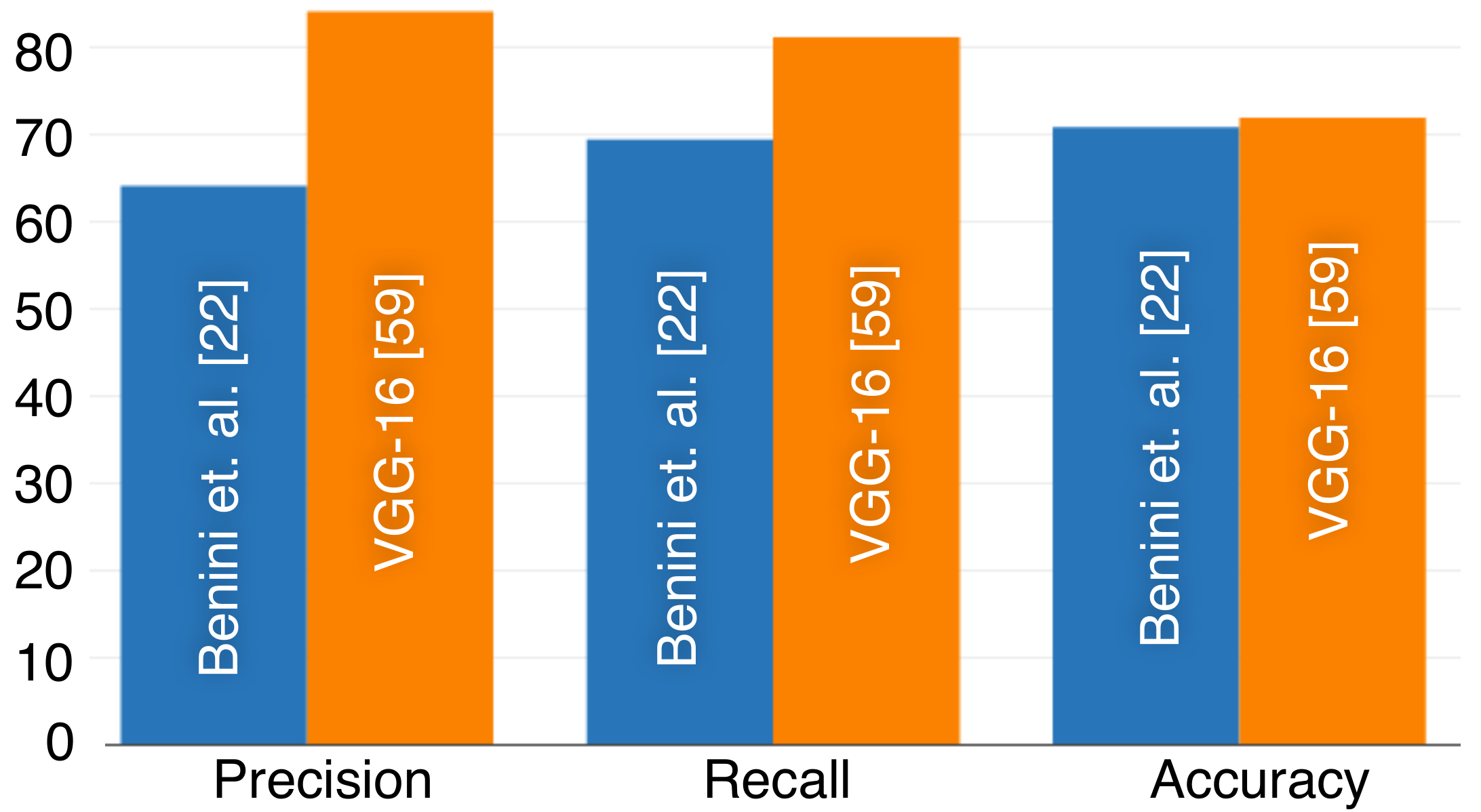}
        \caption{Antonioni's films}
      \end{subfigure}
      %\par
      \begin{subfigure}{0.48\columnwidth}
        \includegraphics[width=\columnwidth]{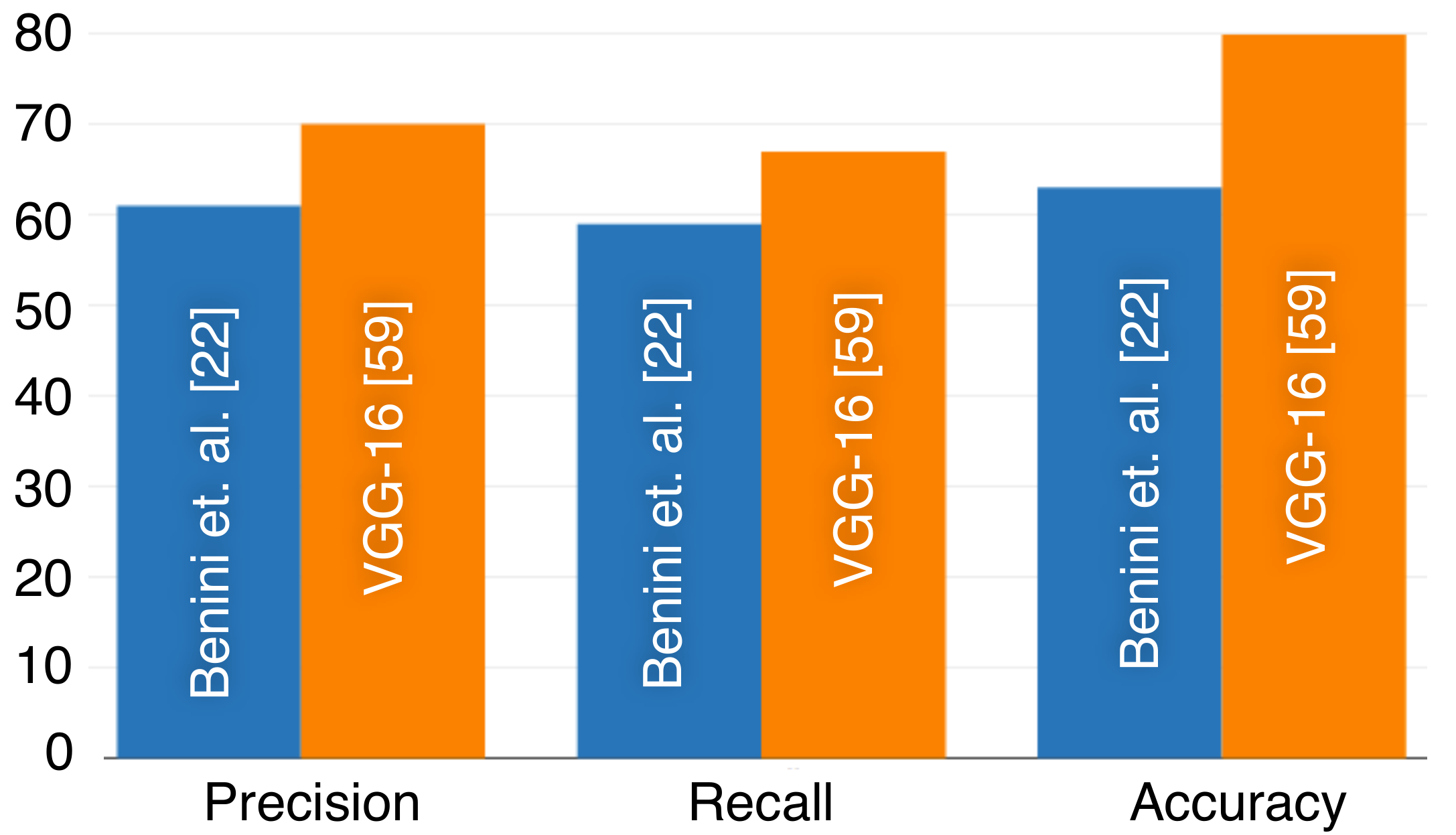}
        \caption{Fellini's films}
      \end{subfigure}
      %\par
\caption{Shot scale classification comparison: VGG-16 \cite{SZ14a} (orange bars) and the method in \cite{BSA16} (blue bars) on 18 movies by (a) Antonioni, and (b) Fellini.}
\label{fig:SDist-comparison}
\end{figure}

To assess the improvement with respect to state-of-the-art, we compare the performance of the best DL architecture (VGG-16) with respect to \cite{BSA16} on the test movies in Antonioni's and Fellini's filmographies, for a total of 18 movies.
Figure~\ref{fig:SDist-comparison} show the compared performance, demonstrating the superiority of the deep network approach.
The test is limited to few films (which account for 200,000 analysed frames) since performing one on the full database would be computationally unfeasible for the method in \cite{BSA16}, which requires a long computational time to extract the hand-crafted features.

%

%%%%%%%%%%%%%%%%%%%%%%%%%%%%%%%%%%%%%%%%%%%%%%%%%%%%%%%%%%%%%%%%
%%%%%%%%%%%%%%%%%%%%%%%%%%%%%%%%%%%%%%%%%%%%%%%%%%%%%%%%%%%%%%%%%
\section{Author recognition by automatic features}
\label{sec:exp2}

\begin{figure}[h]
\centering
	\includegraphics[width=0.65\columnwidth]{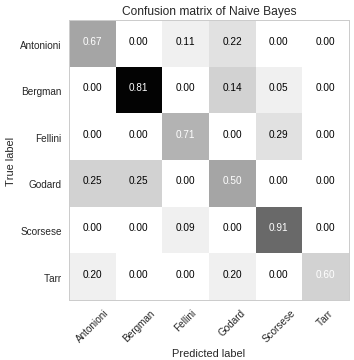}
\caption{Authorship by automatic features: confusion matrix.}
\label{fig:clust_perform_theory_CM-autom}
\end{figure}

To answer Q3 (\emph{is it possible to robustly determine the authorship of a movie using automatically computed shot features?}) we repeat the experiments of Section~\ref{sec:exp1} with automatically extracted features.
Performance are evaluated on the 65 movies used for testing in Section~\ref{ssec:exp-scale}, thus excluding the 55 used for training the CNN.
As done before we employ a leave-one-out of cross-validation, and compute precision, recall, and accuracy, exploiting the Gaussian N\"aive Bayes classifier, as in Section~\ref{sec:exp1}.
Results are presented for every author in Table~\ref{tab:clust_perform_theory-autom} and by the confusion matrix in Figure~\ref{fig:clust_perform_theory_CM-autom}.

\begin{table}[ht]
\centering
      \caption{Authorship recognition based on automatic features. Differences with respect to Table~\ref{tab:clust_perform_theory} are given.}
 \begin{tabular}{ l | c |  c | c |  c  }
    \hline
    \textbf{author} & \textbf{precision} &   \textbf{recall}  & \textbf{f1-score} &   \textbf{movies}\\ \hline \hline
     Antonioni    &   0.60  &    0.67    &  0.63    &   9 \\ \hline
     Bergman    &   0.85   &   0.81  &    0.83   &   21 \\ \hline
     Fellini   &    0.71   &   0.71  &    0.71   &    7 \\ \hline
    Godard   &    0.50    &  0.50   &   0.50    &    12\\ \hline
    Scorsese   &   0.77   &   0.91  &    0.83   &   11\\ \hline
     Tarr    &  1.00  &    0.60    &  0.75   &     5\\ \hline \hline 
\textbf{avg / total}  &     \textbf{0.73}   &   \textbf{0.72}   &   \textbf{0.72}  &    \textbf{65}\\ \hline
\textbf{(difference)}			 &     \textbf{(-0.01)}   &   \textbf{(+0.01)}   &   \textbf{(+0.01)}  &    \textbf{(-12)}\\ \hline
  \end{tabular}
    \label{tab:clust_perform_theory-autom}
  \end{table}

Quite surprisingly there is no drop of performance in recognizing the author with respect to the classification based on GT shot features presented in Table~\ref{tab:clust_perform_theory}, even though tests are conducted on a different number of movies (65 instead of 77).
This implies that automatically computed features are in general robustly extracted.
Using distributions and transition matrices as features has the important advantage that false positive and false negative for each class tend to compensate, de-facto softening the impact of errors on overall classification accuracy.
Therefore even in case of slight differences with respect to ground truth ones, still shot feature distributions and transition patterns change from author to author in a way that is strongly related to the individual style of the director.
%
%%%%%%%%%%%%%%%%%%%%%%%%%%%%%%%%%%%%%%%%%%%%%%%%%%%%%%%%%%%%
%
%%%%%%%%%%%%%%%%%%%%%%%%%%%%%%%%%%%%%%%%%%%%%%%%%%%%%%%
\section{Discussion}
\label{sec:discuss}

We think about works of art as consciously designed and carefully executed to elicit meanings and emotions. 
Therefore features which characterize an artistic form should be meaningful both conceptually and emotionally.
For example, it is true that shot scales are meticulously planned one by one in a film. 
But a scale distribution and sequential pattern in a whole film is something that would not occur to any filmmaker to even think about, let alone plan. 
Here what Antonioni says:
\begin{quote}
\emph{``When I am shooting a film I never think of how I want to shoot something; I simply shoot it. My technique [...] is wholly instinctive and never based on a priori considerations \cite{S69}.''}
\end{quote}

Yet, this study shows that statistical patterns of low-level features are systematically capable of distinguishing between individual authorial styles. 
Even though these patterns cannot be conceived as consciously designed, they underlie intentionally crafted features. 
Whether this relationship is correlational (e.g. narrative features involve specific visual techniques) or perceptual and aesthetic (e.g. a certain aesthetic effect is obtained at a certain frequency of a technical feature value, like eliciting empathy with characters is obtained by frequently shifting from any shot scale to close-up) needs further investigation, stylistic, psychological and statistical. 
%Finding meaningful contexts for these statistical data will tell if these patterns are stylistic in a meaningful sense or merely descriptive and technical.

It is important to notice that sequence patterns are at least as characteristic of individual styles as their distributions. 
We think that the reason is that sequence patterns characterize the dynamics of film form, which has a closer relationship to narration than overall distributions. 
The sequences of shot duration and shot scale might be closely related with such fundamental aesthetic patterns as rhythm, regulation of arousal and of emotional involvement. 
Since the dynamics of these aesthetic effects are the key to regulate dramatic tension in films, which in turn is the key to keep the viewer's attention, it seems probable that sequence patterns are more under conscious authorial control than overall values of the same low-level features. 
However individual styles show a big variety in terms of which feature is the most characteristic of them as compared to the average values across all authors. 
And sometimes no single feature is characteristic for an author's work as with Bergman and Godard. 
However this does not affect the recall rate of their films, which suggests that taking these dimensions together is very robust in the attribution task. 

Despite the large database of 120 movies and the consequent huge number of automatically analyzed frames (more than $772,000$) we are aware that a limitation is the small quantity of the considered directors.
The reduced number of artists might have a strong influence on the performance, which suggests that a further improvement should be directed towards expanding the director set. 
Howerever this work has to be considered as a positive feasibility study on performing automatic attribution of the authorship by means of formal low-level features.
The fact that by using only shot duration and scale it is possible to distinguish among six different authors, with good accuracy, suggests that such analysis can easily scale up to more directors and bigger movie databases.
We are confident that adding other low-level formal features, both perceptual and technical (e.g. depth of field, camera motion, angle, etc.), would increase the accuracy of identification.
% of individual styles.
%, which eventually result in a catalogue of relevant features informing the perception of film style.

Feature learning is nowadays a valid alternative to the use of hand-crafted features.
However learning features directly from data \cite{BCV13} is much more intensive because it needs an enormous number of examples to isolate the distinctive features. 
For example, to discover artists' characteristic visual features, the CNN in \cite{NHP15} works on the Rijksmuseum Challenge database \cite{MG14} which consists of more than 110,000 digitalized artworks by more than 6,000 painters.
Even if we do not exclude to adopt CNN for direct authorship attribution in the future, to operate on movie databases bigger than the current one will require important computational resources.

Conversely, working with formal features has the non trivial advantage of allowing direct testing of hypothesis coming from cinema studies and psychology, thus broadening the scope and the interest of this work to more disciplines beyond video analysis itself.
This work itself is cross-disciplinarily inspired from studies in cinema about the peculiar use of formal features by different directors \cite{Ko14}, and it emotional effects on the viewers \cite{BKP16}.
It is also worth to investigate the effect of such features on viewers from the neuroscience perspective, expanding the work on audio-visual features reconstruction done in \cite{Neuroimagesubmission} and linking it with empathy-related processes felt during the cinematic experience \cite{RWJ12}.

Despite the months spent in performing second-by-second manual annotations, another limitation concerns the labelling of art movies.
During the whole duration of the work, and also after having performed our experiments, we discovered that sometimes the second-based annotations regarding the shot scale were partially inconsistent with processed data.
Although it is not proven that this influences the actual attribution of authorship, still this creates uncertainty about the shot scale recognition performance.

Last, our findings on classification suggest that, for shot scale, using a subdivision in three classes limits the ability of author attribution.
Therefore a possible extension of this work should aim at obtaining a finer automatic classification, able to distinguish beyond CS, MS, and LS, also Extreme Close-up (ECU), Medium Close-up (MCU), Medium Long shot (MLS) and Extreme Long shot (ELS).

%%%%%%%%%%%%%%%%%%%%%%%%%%%%%%%%%%%%%%%%%%%%%%%%%%%%%%%

\section{Conclusion}
\label{sec:conclusion}
In this work we first assess to what extent it is possible to identify the individual styles of movie directors by a statistical analysis of a restricted set of formal low-level features.
We then propose an automatic system able to attribute unseen movies to the correct director by exploiting the information contained in the distribution and sequence patterns of shot duration and shot scale.
Regarding shot scale, we here propose a method for its recognition based on deep neural networks which outperforms existing state-of-the-art.
The experiments are carried out using 120 films of different \'epoques coming from 6 different authors whose styles are consensually considered highly unique and distinguishable in film historiography of modern author cinema.
Findings open up interesting lines of multi-disciplinary research across video analysis, cinema studies, psychology and neurosciences.

\section{Acknowlegments}
\label{sec:ack}
Authors M. Svanera, M. Savardi, and S. Benini equally contributed to this research work. 

\appendices
\section{Movie list}
\label{app-A}

Films with ID\cite{T09} are those with GT on both shot scale and duration. Those with * are used for testing.

\begin{table}[ht]
      \caption{Movies by Michelangelo Antonioni.}
      \centering
 \begin{tabular}{ p{0.6cm} | p{6.2cm} | l  }
    \hline
    \hline
\textbf{Year}  & \textbf{Original title} & \textbf{ID \cite{T09}} \\ 
\hline
1950  & Cronaca di un Amore* & 19651 \\
1953  & I Vinti &  \\ 
1953  & La Signora Senza Camelie*  & 7960 \\ 
1955  & Le Amiche* & 19636 \\ 
1957  & Il Grido * & 19644 \\ 
1960  & L'Avventura* & 19787 \\ 
 1961 & L'Eclisse* & 19647 \\ 
 1962  & La Notte & 19654 \\ 
 1964  & Il Deserto Rosso* & 19632 \\ 
 1966 & Blowup & 19796 \\ 
 1970  & Zabriskie Point & 19649 \\ 
 1975 & Professione: Reporter* & 19659 \\ 
 1980 & Il Mistero di Oberwald &  \\ 
 1982  & Identificazione di una Donna* & 767 \\
  \end{tabular}
    \label{tab:antonioni}
  \end{table}

\begin{table}[ht]
      \caption{Movies by Ingmar Bergman.}
      \centering
 \begin{tabular}{ p{0.6cm} | p{6.2cm} | l  }
    \hline
    \hline
\textbf{Year}  & \textbf{Original title} & \textbf{ID \cite{T09}} \\ 
\hline
 1949  & Fängelse* & 543 \\ 
  1950 & Till Glädje* & 561 \\ 
  1951  & Sommarlek* & 21156 \\ 
  1952  & Kvinnors Väntan* & 1259 \\ 
  1953  & Gycklarnas Afton* & 21152 \\ 
  1953  & Sommaren med Monika* & 21151 \\ 
  1954 & En Lektion i Kärlek* & 5154 \\ 
  1955 & Kvinnordröm &  \\ 
  1955 & Sommarnattens Leende* & 8429 \\ 
  1957  & Det Sjunde Inseglet & 18626 \\ 
  1957 & Smultronstället* & 19755 \\ 
  1958 & Ansiktet &  \\ 
  1958  & Nära Livet &  \\ 
  1960  & Djävulens öga* & 4459 \\ 
  1960 & Jungfrukällan* & 19851 \\ 
  1961 & Sasom i en Spegel* & 7760 \\ 
  1963 & Nattvardsgästerna* & 21155 \\ 
  1963 & Tystnaden* & 7759 \\ 
  1966 & Persona* & 1128 \\ 
 1968 & Skammen* & 19893 \\ 
 1968 & Vargtimmen &  \\ 
  1969  & En Passion* & 2565 \\ 
  1969  & Riten &  \\ 
  1971  & Beröringen &  \\ 
  1972 & Viskningar och Rop* & 10262 \\ 
  1973 & Scener ur ett äktenskap* &  \\ 
  1976  & Ansikte mot Ansikte &  \\ 
  1977 & Das Schlangenei &  \\ 
  1978 & Höstsonaten* & 1144 \\ 
  1980  & Aus dem Leben der Marionetten &  \\ 
  1982  & Fanny och Alexander &  \\ 
  2003 & Saraband* & 5912 \\
 
  \end{tabular}
    \label{tab:bergman}
  \end{table}

  \begin{table}[ht]
      \caption{Movies by Federico Fellini.}
      \centering
 \begin{tabular}{ p{0.6cm} | p{6.2cm} | l  }
    \hline
    \hline
\textbf{Year}  & \textbf{Original title} & \textbf{ID \cite{T09}} \\ 
\hline
   1950 &  Luci del varietà* & 4911 \\ 
  1952 & Lo Sceicco Bianco & 4519 \\ 
  1953 & I Vitelloni & 2846 \\ 
  1954 &La Strada & 12134 \\ 
  1955 & Il Bidone* & 4732 \\ 
  1957  & Le notti di Cabiria & 11020 \\ 
  1960 & La Dolce Vita* & \\
  1963 & 8\sfrac{1}{2} & 5753 \\ 
  1965 & Giulietta degli Spiriti* & 10969 \\ 
  1969 & Satyricon* & 19606 \\ 
  1972 & Roma &  \\ 
  1973 & Amarcord* & 1388 \\ 
  1976 & Il Casanova di Federico Fellini &  \\ 
  1978 & Prova d'Orchestra &  \\ 
  1980  & La Città delle Donne &  \\ 
  1983 & E la Nave Va* & 4548 \\ 
  1986 & Ginger and Fred &  \\ 
  1990 & La Voce della Luna & \\
   \end{tabular}
    \label{tab:fellini}
  \end{table}

    \begin{table}[ht]
      \caption{Movies by Jean-Luc Godard.}
      \centering
 \begin{tabular}{ p{0.6cm} | p{6.2cm} | l  }
    \hline
    \hline
\textbf{Year}  & \textbf{Original title} & \textbf{ID \cite{T09}} \\ 
\hline
 1960 & À bout de souffle* & 10264 \\ 
  1960 & Le Petit Soldat  &  \\ 
  1961  & Une Femme est une Femme* & 3863 \\ 
  1962 & Vivre Sa Vie  & 17623 \\ 
  1963 & Le Mépris* & 12265 \\ 
  1963 & Les Carabiniers* & 606 \\ 
  1964 & Bande à Part* & 18869 \\ 
  1964 & Une Femme Mariée & 616 \\ 
  1965 & Alphaville: une étrange Aventure de Lemmy Caution* & 19533 \\ 
  1965 & Pierrot le Fou & 617 \\ 
  1966 & Made in USA* & 3676 \\ 
  1966 & Masculin Féminin* & 10141 \\ 
  1967 & 2 ou 3 Choses Que Je Sais d'Elle &  \\ 
  1967 & La Chinoise* & 609 \\ 
  1967 & Weekend* &  \\ 
  1969 & Le Gai Savoir &  \\ 
  1972 & Tout va Bien & 3287 \\ 
  1980 & Sauve qui Peut la Vie &  \\ 
  1983 & Prénom Carmen* & 11903 \\ 
  1985 & Détective &  \\ 
  1985 & Je Vous Salue Marie &  \\ 
  1987 & King Lear &  \\ 
  1987 & Soigne ta Droite &  \\ 
  1990 & Nouvelle Vague &  \\ 
  1991 & Allemagne Année 90 Neuf Zéro &  \\ 
  1996 & For Ever Mozart &  \\ 
  2001 & Éloge de l'Amour &  \\ 
  2004 & Notre Musique &  \\ 
  2010 & Film Socialisme* & 11902 \\ 
   \end{tabular}
    \label{tab:godard}
  \end{table}
  
   \begin{table}[ht]
      \caption{Movies by Martin Scorsese.}
      \centering
 \begin{tabular}{ p{0.6cm} | p{6.2cm} | l  }
    \hline
    \hline
\textbf{Year}  & \textbf{Original title} & \textbf{ID \cite{T09}} \\ 
\hline
1967  & Who's that Knocking at my Door &  \\ 
 1972  & Boxcar Bertha &  \\ 
 1973  & Mean Streets* & 12301 \\ 
  1974  & Alice Doesn't Live Here Anymore &  \\ 
  1976 & Taxi Driver* & 3462 \\ 
  1977 & New York, New York &  \\ 
  1980 & Raging Bull* & 3477 \\ 
  1983 & The King of Comedy &  \\ 
  1985 & After Hours* & 13314 \\ 
  1986 & The Color of Money &  \\ 
  1988 & The Last Temptation of Christ &  \\ 
  1990 & Goodfellas* & 13212 \\ 
  1991 & Cape Fear &  \\ 
  1993  & The Age of Innocence &  \\ 
  1995 & Casino* & 6817 \\ 
  1997 & Kundun &  \\ 
  1999 & Bringing Out the Dead &  \\ 
  2002 & Gangs of New York* & 7118 \\ 
  2004 & The Aviator* & 13674 \\ 
  2006 & The Departed* & 1574 \\ 
  2010 & Shutter Island* & 7009 \\ 
  2011 & Hugo* & 16418 \\ 
  2013 & The Wolf of Wall Street & \\
   \end{tabular}
    \label{tab:scorsese}
  \end{table}

    \begin{table}[ht]
      \caption{Movies by Béla Tarr.}
\centering
 \begin{tabular}{ p{0.6cm} | p{6.2cm} | l  }
    \hline
    \hline
\textbf{Year}  & \textbf{Original title} & \textbf{ID \cite{T09}} \\ 
\hline
1977  & Családi t\H{u}zfészek* & 2756 \\ 
  1981  & Szabadgyalog &  \\ 
  1982   & Macbeth  &  \\ 
  1982  & Panelkapcsolat  &  \\ 
  1985 & \H{O}szi Almanach* & 3453 \\ 
  1988 & Kárhozat* & 10656 \\ 
  1994 & Sátántangó & 801 \\ 
  2000 & Werckmeister Harmóniák* & 2644 \\ 
  2007 & A Londoni Férfi* & 12376 \\ 
  2011 & A Torinói Ló* & 9760 \\

   \end{tabular}
    \label{tab:tarr}
  \end{table}

\bibliographystyle{IEEEtran}
\bibliography{IEEEabrv,bibliography.bib}

\begin{thebibliography}{10}
\providecommand{\url}[1]{#1}
\csname url@samestyle\endcsname
\providecommand{\newblock}{\relax}
\providecommand{\bibinfo}[2]{#2}
\providecommand{\BIBentrySTDinterwordspacing}{\spaceskip=0pt\relax}
\providecommand{\BIBentryALTinterwordstretchfactor}{4}
\providecommand{\BIBentryALTinterwordspacing}{\spaceskip=\fontdimen2\font plus
\BIBentryALTinterwordstretchfactor\fontdimen3\font minus
  \fontdimen4\font\relax}
\providecommand{\BIBforeignlanguage}[2]{{%
\expandafter\ifx\csname l@#1\endcsname\relax
\typeout{** WARNING: IEEEtran.bst: No hyphenation pattern has been}%
\typeout{** loaded for the language `#1'. Using the pattern for}%
\typeout{** the default language instead.}%
\else
\language=\csname l@#1\endcsname
\fi
#2}}
\providecommand{\BIBdecl}{\relax}
\BIBdecl

\bibitem{S-06}
\BIBentryALTinterwordspacing
B.~Salt, \emph{Moving Into Pictures: More on Film History, Style, and
  Analysis}.\hskip 1em plus 0.5em minus 0.4em\relax Starword, 2006. [Online].
  Available: \url{http://books.google.it/books?id=wW8cAgAACAAJ}
\BIBentrySTDinterwordspacing

\bibitem{buzzfeed}
\BIBentryALTinterwordspacing
``The signature trademarks of 14 famous directors,'' online; accessed:
  18-May-2017. [Online]. Available:
  \url{https://www.buzzfeed.com/mrloganrhoades/the-signature-trademarks-of-14-famous-directors?utm_term=.cje9yQpvE#.nbmbXq1GD}
\BIBentrySTDinterwordspacing

\bibitem{Ko14}
A.~B. Kov{\'a}cs, ``Shot scale distribution: an authorial fingerprint or a
  cognitive pattern?'' \emph{Projections}, vol.~8, no.~2, 2014.

\bibitem{W40}
C.~B. Williams, ``A note on the statistical analysis of sentence-length as a
  criterion of literary style,'' \emph{Biometrika}, vol.~31, no. 3/4, pp.
  356--361, 1940.

\bibitem{MOVY01}
B.~S. Manjunath, J.-R. Ohm, V.~V. Vasudevan, and A.~Yamada, ``Color and texture
  descriptors,'' \emph{IEEE Transactions on Circuits and Systems for Video
  Technology}, vol.~11, no.~6, pp. 703--715, 2001.

\bibitem{VM94}
P.~Valdez and A.~Mehrabian, ``Effects of color on emotions,'' \emph{Journal of
  Experimental Psychology}, vol. 123, no.~4, pp. 394--409, 1994.

\bibitem{WC06}
H.~L. Wang and L.~F. Cheong, ``Affective understanding in film,'' \emph{IEEE
  Transactions on Circuits and Systems for Video Technology}, vol.~16, no.~6,
  pp. 689--704, June 2006.

\bibitem{WGG07}
J.~V. de~Weijer, T.~Gevers, and A.~Gijsenij, ``Edge-based color constancy,''
  \emph{IEEE Transactions on Image Processing}, vol.~16, no.~9, pp. 2207--2214,
  Sep. 2007.

\bibitem{CBML09}
L.~Canini, S.~Benini, P.~Migliorati, and R.~Leonardi, ``Emotional identity of
  movies,'' in \emph{Proceedings of the 16th IEEE International Conference on
  Image Processing}, Cairo, Egypt, 7-11 November 2009.

\bibitem{TSK00}
Y.-P. Tan, D.~D. Saur, S.~R. Kulkarni, and P.~J. Ramadge, ``Rapid estimation of
  camera motion from compressed video with application to video annotation,''
  \emph{IEEE Trans. on Circuits and Systems for Video Technology}, vol.~10, pp.
  133--146, 2000.

\bibitem{EST04}
R.~Ewerth, M.~Schwalb, P.~Tessmann, and B.~Freisleben, ``Estimation of
  arbitrary camera motion in mpeg videos,'' in \emph{Proceedings of the 17th
  International Conference on Pattern Recognition, 2004. ICPR 2004.}, vol.~1,
  Aug 2004, pp. 512--515 Vol.1.

\bibitem{BMS14}
S.~Bhattacharya, R.~Mehran, R.~Sukthankar, and M.~Shah, ``Classification of
  cinematographic shots using lie algebra and its application to complex event
  recognition,'' \emph{Multimedia, IEEE Transactions on}, vol.~16, no.~3, pp.
  686--696, 4 2014.

\bibitem{ZGZ-02}
W.~Zeng, W.~Gao, and D.~Zhao, ``Video indexing by motion activity maps,'' in
  \emph{Image Processing. 2002. Proceedings. 2002 International Conference on},
  vol.~1.\hskip 1em plus 0.5em minus 0.4em\relax IEEE, 2002, pp. I--912.

\bibitem{JD-01}
S.~Jeannin and A.~Divakaran, ``Mpeg-7 visual motion descriptors,''
  \emph{Circuits and Systems for Video Technology, IEEE Transactions on},
  vol.~11, no.~6, pp. 720--724, 2001.

\bibitem{RSS05}
Z.~Rasheed, Y.~Sheikh, and M.~Shah, ``On the use of computable features for
  film classification,'' \emph{IEEE Transactions on Circuits and Systems for
  Video Technology}, vol.~15, no.~1, pp. 52--64, Jan 2005.

\bibitem{ZS11}
\BIBentryALTinterwordspacing
S.~Zhuo and T.~Sim, ``Defocus map estimation from a single image,''
  \emph{Pattern Recogn.}, vol.~44, no.~9, pp. 1852--1858, Sep. 2011. [Online].
  Available: \url{http://dx.doi.org/10.1016/j.patcog.2011.03.009}
\BIBentrySTDinterwordspacing

\bibitem{H02}
A.~Hanjalic, ``Shot-boundary detection: unraveled and resolved?'' \emph{IEEE
  Transactions on Circuits and Systems for Video Technology}, vol.~12, no.~2,
  pp. 90--105, Feb 2002.

\bibitem{SOD10}
\BIBentryALTinterwordspacing
A.~F. Smeaton, P.~Over, and A.~R. Doherty, ``Video shot boundary detection:
  Seven years of trecvid activity,'' \emph{Comput. Vis. Image Underst.}, vol.
  114, no.~4, pp. 411--418, Apr. 2010. [Online]. Available:
  \url{http://dx.doi.org/10.1016/j.cviu.2009.03.011}
\BIBentrySTDinterwordspacing

\bibitem{Choros09}
K.~Choro{\'s}, ``Video shot selection and content-based scene detection for
  automatic classification of tv sports news,'' in \emph{Internet -- Technical
  Development and Applications}, ser. Advances in Intelligent and Soft
  Computing.\hskip 1em plus 0.5em minus 0.4em\relax Springer Berlin /
  Heidelberg, 2009, vol.~64, pp. 73--80.

\bibitem{WC09}
H.~L. Wang and L.-F. Cheong, ``Taxonomy of directing semantics for film shot
  classification,'' \emph{Circuits and Systems for Video Technology, IEEE
  Transactions on}, vol.~19, no.~10, pp. 1529--1542, 10 2009.

\bibitem{A-91}
D.~Arijon, \emph{Grammar of the Film Language}.\hskip 1em plus 0.5em minus
  0.4em\relax Silman-James Press, September 1991.

\bibitem{BSA16}
S.~Benini, M.~Svanera, N.~Adami, R.~Leonardi, and A.~B. Kov{\'a}cs, ``Shot
  scale distribution in art films,'' \emph{Multimedia Tools and Applications},
  pp. 1--29, 2016.

\bibitem{CBL11}
L.~Canini, S.~Benini, and R.~Leonardi, ``Affective analysis on patterns of shot
  types in movies,'' in \emph{Image and Signal Processing and Analysis (ISPA),
  2011 7th Int. Symposium on}.\hskip 1em plus 0.5em minus 0.4em\relax IEEE,
  2011, pp. 253--258.

\bibitem{SW05}
\BIBentryALTinterwordspacing
C.~G. Snoek and M.~Worring, ``Multimodal video indexing: A review of the
  state-of-the-art,'' \emph{Multimedia Tools and Applications}, vol.~25, no.~1,
  pp. 5--35, 2005. [Online]. Available:
  \url{http://dx.doi.org/10.1023/B:MTAP.0000046380.27575.a5}
\BIBentrySTDinterwordspacing

\bibitem{NH01}
H.~R. Naphide and T.~S. Huang, ``A probabilistic framework for semantic video
  indexing, filtering, and retrieval,'' \emph{IEEE Transactions on Multimedia},
  vol.~3, no.~1, pp. 141--151, Mar 2001.

\bibitem{HX05}
A.~Hanjalic and L.-Q. Xu, ``Affective video content representation and
  modeling,'' \emph{IEEE Transactions on Multimedia}, vol.~7, no.~1, pp.
  143--154, Feb 2005.

\bibitem{BCL11}
S.~Benini, L.~Canini, and R.~Leonardi, ``A connotative space for supporting
  movie affective recommendation,'' \emph{IEEE Transactions on Multimedia},
  vol.~13, no.~6, pp. 1356--1370, Dec 2011.

\bibitem{CBL13a}
L.~Canini, S.~Benini, and R.~Leonardi, ``Affective recommendation of movies
  based on selected connotative features,'' \emph{IEEE Trans. on Circuits \&
  Systems for Video Technology}, vol.~23, no.~4, pp. 636--647, April 2013.

\bibitem{NMZ05}
C.-W. Ngo, Y.-F. Ma, and H.-J. Zhang, ``Video summarization and scene detection
  by graph modeling,'' \emph{IEEE Transactions on Circuits and Systems for
  Video Technology}, vol.~15, no.~2, pp. 296--305, Feb 2005.

\bibitem{MHL05}
Y.-F. Ma, X.-S. Hua, L.~Lu, and H.-J. Zhang, ``A generic framework of user
  attention model and its application in video summarization,'' \emph{IEEE
  Transactions on Multimedia}, vol.~7, no.~5, pp. 907--919, Oct 2005.

\bibitem{ZHK10}
\BIBentryALTinterwordspacing
H.~Zhou, T.~Hermans, A.~V. Karandikar, and J.~M. Rehg, ``Movie genre
  classification via scene categorization,'' in \emph{Proceedings of the 18th
  ACM International Conference on Multimedia}, ser. MM '10.\hskip 1em plus
  0.5em minus 0.4em\relax New York, NY, USA: ACM, 2010, pp. 747--750. [Online].
  Available: \url{http://doi.acm.org/10.1145/1873951.1874068}
\BIBentrySTDinterwordspacing

\bibitem{MYL11}
\BIBentryALTinterwordspacing
T.~Mei, B.~Yang, X.-S. Hua, and S.~Li, ``Contextual video recommendation by
  multimodal relevance and user feedback,'' \emph{ACM Trans. Inf. Syst.},
  vol.~29, no.~2, pp. 10:1--10:24, Apr. 2011. [Online]. Available:
  \url{http://doi.acm.org/10.1145/1961209.1961213}
\BIBentrySTDinterwordspacing

\bibitem{DH00}
\BIBentryALTinterwordspacing
S.~B. Dermer and J.~B. Hutchings, ``Utilizing movies in family therapy:
  Applications for individuals, couples, and families,'' \emph{The American
  Journal of Family Therapy}, vol.~28, no.~2, pp. 163--180, 2000. [Online].
  Available: \url{http://dx.doi.org/10.1080/019261800261734}
\BIBentrySTDinterwordspacing

\bibitem{DCC11}
\BIBentryALTinterwordspacing
M.~B. Devlin, L.~T. Chambers, and C.~Callison, ``Targeting mood: Using comedy
  or serious movie trailers,'' \emph{Journal of Broadcasting \& Electronic
  Media}, vol.~55, no.~4, pp. 581--595, 2011. [Online]. Available:
  \url{http://dx.doi.org/10.1080/08838151.2011.620668}
\BIBentrySTDinterwordspacing

\bibitem{NHP15}
N.~van Noord, E.~Hendriks, and E.~Postma, ``Toward discovery of the artist's
  style: Learning to recognize artists by their artworks,'' \emph{IEEE Signal
  Processing Magazine}, vol.~32, no.~4, pp. 46--54, July 2015.

\bibitem{GEB15}
\BIBentryALTinterwordspacing
L.~A. Gatys, A.~S. Ecker, and M.~Bethge, ``A neural algorithm of artistic
  style,'' \emph{arXiv}, Aug 2015. [Online]. Available:
  \url{http://arxiv.org/abs/1508.06576}
\BIBentrySTDinterwordspacing

\bibitem{wiki-art-film}
\BIBentryALTinterwordspacing
Wikipedia, ``Art film --- wikipedia{,} the free encyclopedia,'' 2015, [Online;
  accessed 20-Mar-2015]. [Online]. Available:
  \url{http://en.wikipedia.org/w/index.php?title=Art_film&oldid=646428976}
\BIBentrySTDinterwordspacing

\bibitem{CBD11}
\BIBentryALTinterwordspacing
J.~E. Cutting, K.~L. Brunick, J.~E. DeLong, C.~Iricinschi, and A.~Candan,
  ``Quicker, faster, darker: Changes in hollywood film over 75 years,''
  \emph{i-Perception}, vol.~2, no.~6, pp. 569--576, 2011. [Online]. Available:
  \url{http://dx.doi.org/10.1068/i0441aap}
\BIBentrySTDinterwordspacing

\bibitem{C16a}
\BIBentryALTinterwordspacing
J.~E. Cutting, ``The evolution of pace in popular movies,'' \emph{Cognitive
  Research: Principles and Implications}, vol.~1, no.~1, p.~30, 2016. [Online].
  Available: \url{http://dx.doi.org/10.1186/s41235-016-0029-0}
\BIBentrySTDinterwordspacing

\bibitem{S12}
T.~J. Smith, ``Attentional theory of cinematic continuity,'' \emph{Projections:
  The Journal for Movies and Mind}, vol.~6, no.~1, pp. 1--27, 2012.

\bibitem{C16}
\BIBentryALTinterwordspacing
J.~E. Cutting, ``Narrative theory and the dynamics of popular movies,''
  \emph{Psychonomic Bulletin {\&} Review}, vol.~23, no.~6, pp. 1713--1743,
  2016. [Online]. Available: \url{http://dx.doi.org/10.3758/s13423-016-1051-4}
\BIBentrySTDinterwordspacing

\bibitem{BKP16}
\BIBentryALTinterwordspacing
K.~B{\'a}lint, T.~Klausch, and T.~P{\'o}lya, ``Watching closely,''
  \emph{Journal of Media Psychology}, vol.~0, no.~0, pp. 1--10, 2016. [Online].
  Available: \url{http://dx.doi.org/10.1027/1864-1105/a000189}
\BIBentrySTDinterwordspacing

\bibitem{SBA15}
\BIBentryALTinterwordspacing
M.~Svanera, S.~Benini, N.~Adami, R.~Leonardi, and A.~B. Kov{\'a}cs,
  ``Over-the-shoulder shot detection in art films,'' in \emph{13th
  International Workshop on Content-Based Multimedia Indexing, {CBMI} 2015,
  Prague, Czech Republic, June 10-12, 2015}.\hskip 1em plus 0.5em minus
  0.4em\relax {IEEE}, 2015, pp. 1--6. [Online]. Available:
  \url{http://dx.doi.org/10.1109/CBMI.2015.7153627}
\BIBentrySTDinterwordspacing

\bibitem{cinemetrics}
\BIBentryALTinterwordspacing
``Cinemetrics: Movie measurement and study tool database,'' 2017, [Online;
  accessed 18-May-'17]. [Online]. Available: \url{www.cinemetrics.lv}
\BIBentrySTDinterwordspacing

\bibitem{T09}
Y.~Tsivian, ``Cinemetrics, part of the humanities's cyberinfrastructure,''
  \emph{Digital Tools in Media Studies}, vol.~9, no. 93-100, p.~94, 2009.

\bibitem{t_SNE}
L.~v.~d. Maaten and G.~Hinton, ``Visualizing data using t-sne,'' \emph{Journal
  of Machine Learning Research}, vol.~9, no. Nov, pp. 2579--2605, 2008.

\bibitem{project}
\BIBentryALTinterwordspacing
``Who is the director of this movie?'' 2017, [Online; accessed 18-May-2017].
  [Online]. Available:
  \url{http://projects.i-ctm.eu/en/project/who-director-movie-shot-analysis-art-films}
\BIBentrySTDinterwordspacing

\bibitem{scikit-learn}
F.~Pedregosa, G.~Varoquaux, A.~Gramfort, V.~Michel, B.~Thirion, O.~Grisel,
  M.~Blondel, P.~Prettenhofer, R.~Weiss, V.~Dubourg, J.~Vanderplas, A.~Passos,
  D.~Cournapeau, M.~Brucher, M.~Perrot, and E.~Duchesnay, ``Scikit-learn:
  Machine learning in {P}ython,'' \emph{Journal of Machine Learning Research},
  vol.~12, pp. 2825--2830, 2011.

\bibitem{Z04}
H.~Zhang, ``The optimality of naive bayes,'' \emph{AA}, vol.~1, no.~2, p.~3,
  2004.

\bibitem{K13}
\BIBentryALTinterwordspacing
A.~B. Kovács, \emph{The Cinema of Béla Tarr: The Circle Closes}.\hskip 1em plus
  0.5em minus 0.4em\relax Columbia University Press, 2013. [Online]. Available:
  \url{http://www.jstor.org/stable/10.7312/kova16530}
\BIBentrySTDinterwordspacing

\bibitem{FB81}
\BIBentryALTinterwordspacing
M.~A. Fischler and R.~C. Bolles, ``Random sample consensus: A paradigm for
  model fitting with applications to image analysis and automated
  cartography,'' \emph{Commun. ACM}, vol.~24, no.~6, pp. 381--395, Jun. 1981.
  [Online]. Available: \url{http://doi.acm.org/10.1145/358669.358692}
\BIBentrySTDinterwordspacing

\bibitem{O15}
\BIBentryALTinterwordspacing
J.~A. Ohlson and S.~Kim, ``Linear valuation without ols: the theil-sen
  estimation approach,'' \emph{Review of Accounting Studies}, vol.~20, no.~1,
  pp. 395--435, Mar 2015. [Online]. Available:
  \url{http://dx.doi.org/10.1007/s11142-014-9300-0}
\BIBentrySTDinterwordspacing

\bibitem{AM14}
E.~Apostolidis and V.~Mezaris, ``Fast shot segmentation combining global and
  local visual descriptors,'' in \emph{2014 IEEE International Conference on
  Acoustics, Speech and Signal Processing (ICASSP)}, May 2014, pp. 6583--6587.

\bibitem{AL99}
N.~Adami and R.~Leonardi, ``Identification of editing effect in image sequences
  by statistical modeling,'' in \emph{in Proc. of the 1999 Picture Coding
  Symposium (PCS 1999)}, 1999, pp. 157--160.

\bibitem{shotcut-code}
\BIBentryALTinterwordspacing
{Multimedia Knowledge and Social Media Analytics Laboratory}, ``Video shot and
  scene segmentation,'' online; accessed: 18-May-2017. [Online]. Available:
  \url{http://mklab.iti.gr/project/video-shot-segm}
\BIBentrySTDinterwordspacing

\bibitem{KSH12}
A.~Krizhevsky, I.~Sutskever, and G.~E. Hinton, ``Imagenet classification with
  deep convolutional neural networks,'' in \emph{Advances in Neural Information
  Processing Systems 25}.\hskip 1em plus 0.5em minus 0.4em\relax Curran
  Associates, Inc., 2012, pp. 1097--1105.

\bibitem{SLJ15}
C.~Szegedy, W.~Liu, Y.~Jia, P.~Sermanet, S.~Reed, D.~Anguelov, D.~Erhan,
  V.~Vanhoucke, and A.~Rabinovich, ``Going deeper with convolutions,'' in
  \emph{2015 IEEE Conference on Computer Vision and Pattern Recognition
  (CVPR)}, June 2015, pp. 1--9.

\bibitem{SZ14a}
\BIBentryALTinterwordspacing
K.~Simonyan and A.~Zisserman, ``Very deep convolutional networks for
  large-scale image recognition,'' \emph{CoRR}, vol. abs/1409.1556, 2014.
  [Online]. Available: \url{http://arxiv.org/abs/1409.1556}
\BIBentrySTDinterwordspacing

\bibitem{imagenet}
J.~Deng, W.~Dong, R.~Socher, L.-J. Li, K.~Li, and L.~Fei-Fei, ``{ImageNet: A
  Large-Scale Hierarchical Image Database},'' in \emph{CVPR09}, 2009.

\bibitem{S69}
\BIBentryALTinterwordspacing
C.~T. Samuels, ``Interview with {M}ichelangelo {A}ntonioni in {R}ome,'' July
  29th 1969, [Online; accessed 11-Jul-2017]. [Online]. Available:
  \url{http://zakka.dk/euroscreenwriters/interviews/michelangelo_antonioni_02.htm}
\BIBentrySTDinterwordspacing

\bibitem{BCV13}
\BIBentryALTinterwordspacing
Y.~Bengio, A.~C. Courville, and P.~Vincent, ``Representation learning: A review
  and new perspectives.'' \emph{IEEE Trans. Pattern Anal. Mach. Intell.},
  vol.~35, no.~8, pp. 1798--1828, 2013. [Online]. Available:
  \url{http://dblp.uni-trier.de/db/journals/pami/pami35.html#BengioCV13}
\BIBentrySTDinterwordspacing

\bibitem{MG14}
T.~Mensink and J.~van Gemert, ``The rijksmuseum challenge: Museum-centered
  visual recognition,'' in \emph{ACM International Conference on Multimedia
  Retrieval (ICMR)}, 2014.

\bibitem{Neuroimagesubmission}
G.~Raz, M.~Svanera, N.~Singer, G.~Gilam, M.~Bleich~Cohen, T.~Lin, R.~Admon,
  T.~Gonen, A.~Thaler, R.~Y. Granot, R.~Goebel, S.~Benini, and G.~Valente,
  ``Robust inter-subject audiovisual decoding in functional magnetic resonance
  imaging ({fMRI}) using multivariate regression,'' \emph{NeuroImage (in
  review)}, Jun. 2017.

\bibitem{RWJ12}
\BIBentryALTinterwordspacing
G.~Raz, Y.~Winetraub, Y.~Jacob, S.~Kinreich, A.~Maron-Katz, G.~Shaham,
  I.~Podlipsky, G.~Gilam, E.~Soreq, and T.~Hendler, ``Portraying emotions at
  their unfolding: {A} multilayered approach for probing dynamics of neural
  networks,'' \emph{NeuroImage}, vol.~60, no.~2, pp. 1448--1461, Apr. 2012.
  [Online]. Available:
  \url{http://www.sciencedirect.com/science/article/pii/S1053811912000250}
\BIBentrySTDinterwordspacing

\end{thebibliography}

%%%%%%%%%%%%%%%%%%%%%%%%%%%%%%%%%%%%%%%%%%%%%%%%%%%%%%

\begin{IEEEbiography}
    [{\includegraphics[width=1in,height=1.25in,clip,keepaspectratio]{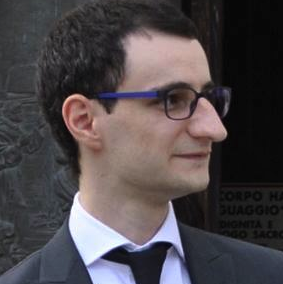}}]{Michele Svanera} obtained his B.Sc, M.Sc, and PhD in Telecommunication Eng. at University of Brescia. During his PhD he moved across AI and Cognitive Neuroscience to understand brain activities under movie stimulus, visiting important brain centres in Maastricht and Tel Aviv. Since May 2017 he joined Lars Muckli's Lab, University of Glasgow, as a postdoc, working on high-field fMRI and localisation of non-feedforward sources in primary visual cortex.
\end{IEEEbiography}

\begin{IEEEbiography}
    [{\includegraphics[width=1in,height=1.25in,clip,keepaspectratio]{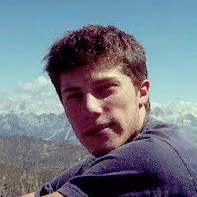}}]{Mattia Savardi} received his M.Sc. degree in Communication Technologies and Multimedia (cum laude) at the University of Brescia in 2016.
He is currently a Ph.D student at the University of Brescia since November 2016. 
His efforts are on semantic and multimedia content analysis and biomedical signal and imaging applications using Computer Vision and Deep Learning techniques.
\end{IEEEbiography}

\begin{IEEEbiography}
    [{\includegraphics[width=1in,height=1.25in,clip,keepaspectratio]{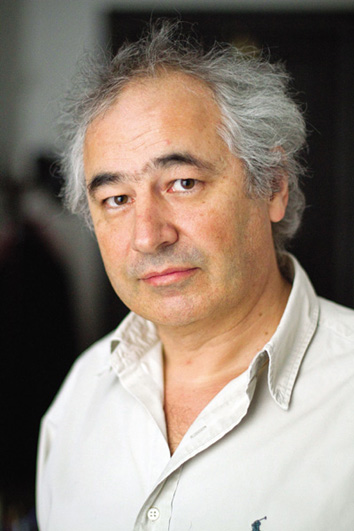}}]{Andra\'s B\'alint Kov\'acs} is professor and founding chair of the Dept. of Film Studies at ELTE University, Budapest. 
He teaches history, analysis and cognitive theory of cinema. His research includes psychological studies of viewer's reactions to causal connections in filmic narratives, psychology of perception of shot scales, and statistical analysis of film style. 
He has been visiting professor at Univ. of Stockholm, Universit\`e de la Nouvelle Sorbonne, \'Ecole Normale Sup\'erieure, and UC San Diego.
 \end{IEEEbiography}

\begin{IEEEbiography}
    [{\includegraphics[width=1in,height=1.25in,clip,keepaspectratio]{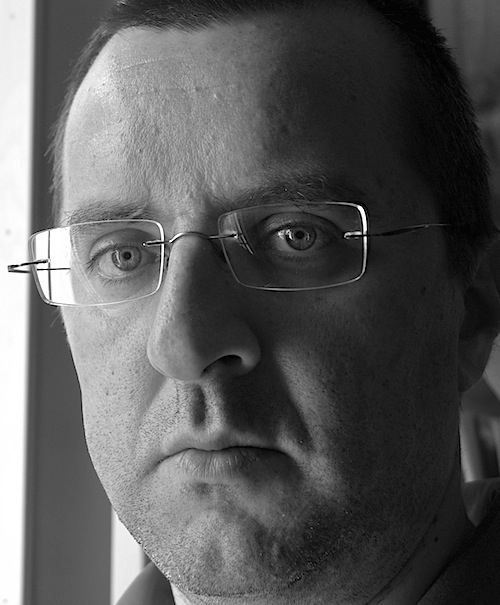}}]{Alberto Signoroni}
(M'03) received M.Sc. in Electronic Engineering 97 and Ph.D. in Information Engineering 01 from the University of Brescia, Italy. Currently he is an Assistant Professor with the Information Engineering Department at the University of Brescia. His research interests include 3D computer vision and geometry processing, biomedical image analysis, multidimensional and hyperspectral image processing. 
\end{IEEEbiography}

\begin{IEEEbiography}
    [{\includegraphics[width=1in,height=1.25in,clip,keepaspectratio]{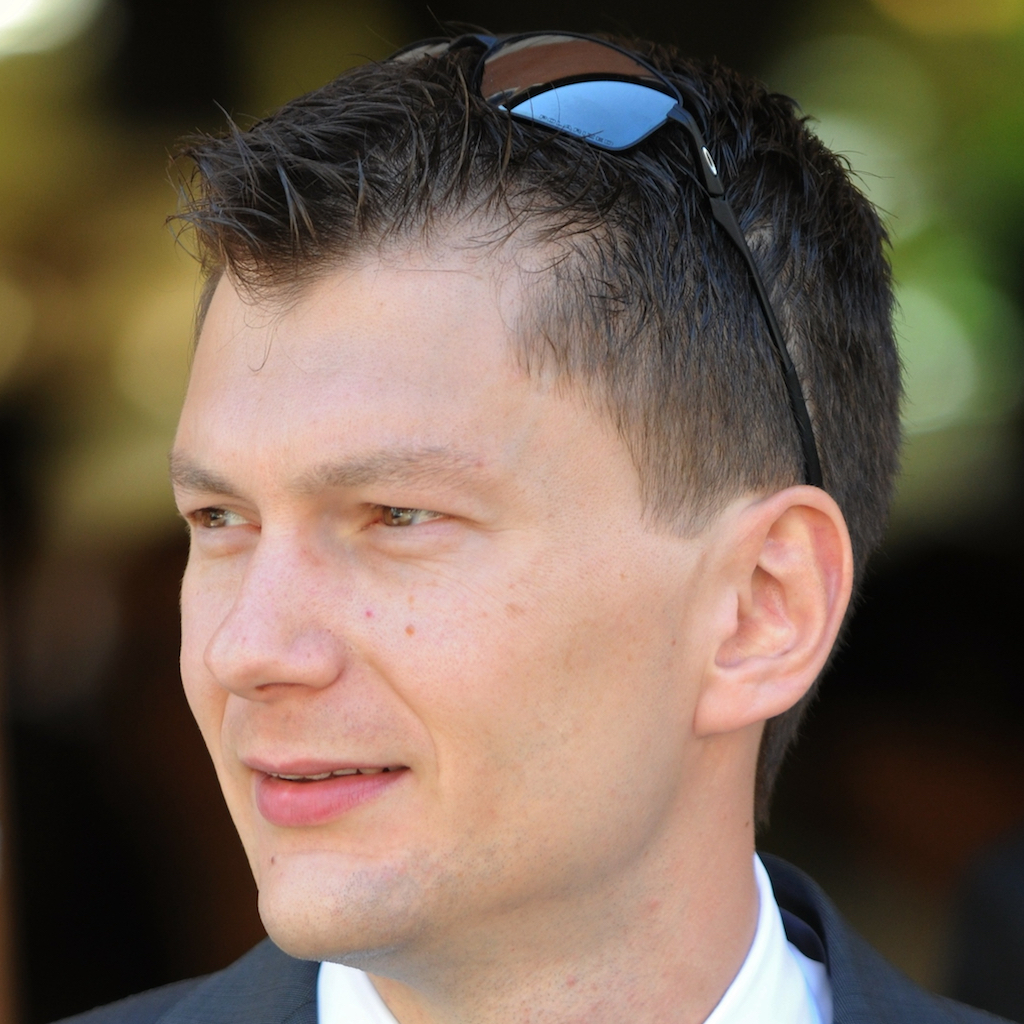}}]{Sergio Benini}
received both his MSc in Electronic Engineering in 2000 (cum laude) and PhD in Information Engineering (2006) from the University of Brescia, IT. 
Between '01 and '03 he was with Siemens Mobile Communications. 
During his Ph.D. he spent one year in British Telecom UK. 
Since 2005 he is Assistant Professor at the University of Brescia.
In '12 he co-founded Yonder \url{http://yonderlabs.com}, a spin-off in NLP, ML, and Cognitive Computing.
\end{IEEEbiography}
\end{document}